\documentclass[fleqn,usenatbib]{mnras}
\usepackage{newtxtext,newtxmath}
\usepackage[T1]{fontenc}
\usepackage{ae,aecompl}

\usepackage{graphicx,caption}
\usepackage{amsmath}
\usepackage{orcidlink}
\usepackage{graphicx}
\usepackage{xspace}
\usepackage{ragged2e}

\usepackage{array}
\newcommand\HI{H\textsc{i}\xspace}
\newcommand\HIMF{H\textsc{i}MF\xspace}
\newcommand\HIMFs{H\textsc{i}MFs\xspace}
\def\Sref#1{Sec.~\ref{#1}\xspace}
\def\Fref#1{Fig.~\ref{#1}\xspace}
\def\Eref#1{Eq.~\eqref{#1}\xspace}
\def\Tref#1{Table~\ref{#1}\xspace}
\def\Aref#1{Appendix~\ref{#1}\xspace}
\usepackage{color}

\title[Joint Optical-HI mock catalogues]{Joint Optical-\HI mock catalogues and prospects for upcoming \HI surveys}

\author[Bharti \& Bagla]{
Sauraj Bharti\orcidlink{0000-0001-9030-3926}
and Jasjeet Singh Bagla\orcidlink{0000-0002-7749-4155}\thanks{E-mail: \href{mailto:jasjeet@iisermohali.ac.in}{jasjeet@iisermohali.ac.in}}
\\
\\
Department of Physical Sciences, IISER Mohali, Knowledge City, Sector 81, SAS Nagar, Punjab 140306, India
}

\date{Accepted XXX; Received YYY; in original form ZZZ}

\pubyear{2026}

\begin{document}
\label{firstpage}
\pagerange{\pageref{firstpage}--\pageref{lastpage}}
\maketitle

%%%%%%%%%%%%%%%%%%%%%%%%%%%%%%%%%%%%%%%%%%%%%%%%%%%%%%%%%%%%%%%%%%%%%%%%
\begin{abstract}
Atomic hydrogen (\HI) regulates star formation as cold gas fuels star formation.  
It represents a key phase in the baryon cycle.
Redshifted $21$~cm line emission serves as a key tracer for investigating \HI in the interstellar medium (ISM) and circumgalactic medium (CGM) in distant galaxies, and enables the study of galaxy evolution.  
Direct detections of \HI are currently limited to $z \leq 0.4$ due to the inherently weak $21$~cm emission line. 
Ongoing and upcoming large radio surveys aim to detect $21$~cm emission from galaxies up to $z \gtrsim 1$.  
We present a novel approach for creating optical-\HI joint mock catalogues for upcoming SKA precursor surveys: MIGHTEE-\HI, LADUMA and WALLABY. 
Incorporation of optical properties along with \HI makes these mock catalogues a powerful tool for making predictions for surveys and provides a benchmark for exploring the \HI science (e.g., conditional \HIMF and optical-to-\HI scaling relations) expected from these surveys. 
As a case study, we show the use of the joint catalogues for predicting the expected outcome of stacking detection for average \HI mass in galaxies that are below the threshold for direct detection. 
We show that combining stacking observations with only the number of direct detections puts a strong constraint on the \HI mass function.  We also show that this can be done separately for parts of the color-magnitude diagram for galaxies, enabling a fine grained study of galaxy evolution.
This may be used to set priors for the full determination of the \HI mass function.
\end{abstract}

%%%%%%%%%%%%%%%%%%%%%%%%%%%%%%%%%%%%%%%%%%%%%%%%%%%%%%%%%%%%%%%%%%%%%%%%

%%%%%%%%%%%%%%%%%%%%%%%%%%%%%%%%%%%%%%%%%%%%%%%%%%%%%%%%%%%%%%%%%%%%%%%%
\maketitle

%%%%%%%%%%%%%%%%%%%%%%%%%%%%%%%%%%%%%%%%%%%%%%%%%%%%%%%%%%%%%%%%%%%%%%%%
\section{Introduction}

Gravitational clustering leads to the formation of large-scale structure in the Universe.  
Much of the mass in the Universe is in the highly over dense network traced by Galaxies, even though these occupy only a tiny fraction of the volume.  
Galaxies are thus the most prominent visible tracers of matter in the Universe. 
Naturally, these have been the primary focus of extra-galactic observational studies.  
The optically visible matter is dominated by stars in galaxies but the formation of the stellar component depends on the availability of cold gas, and its evolution depends on the interplay of accretion on to halos, cooling of hot gas and formation of atomic and molecular components of the inter-stellar medium, star formation and feedback.  
The combination and interplay of these processes is often referred to as the baryonic cycle \citep[e.g.,][]{Ford2014, Tumlinson2017}. 
It is the gas cycle between the outer circumgalactic medium~(CGM) and the interstellar medium~(ISM) through complex outflows, inflows of cold \HI gas, H$_2$ gas, and metals. 
Neutral Hydrogen (\HI) gas serves as the primary reservoir for the fuel for star formation and it is the key ingredient that drives galaxy evolution. 
Our understanding of galaxy evolution has been improved with the several \HI and optical surveys such as HI Parkes All Sky Survey~\citep[HIPASS;][]{Zwaan_2005}, Arecibo Legacy Fast Arecibo L-band Feed Array \citep[ALFALFA;][]{Giovanelli_2005,Jones,Haynes}, Sloan Digital Sky Surveys \citep[SDSS;][]{Abazajian}, and  Cosmic Evolution Survey~\citep[COSMOS;][]{Scoville_2007}. 
In several studies, it is found that the \HI mass of local galaxies is correlated with their B band optical luminosity and size \citet{Chowdhury2021,Chowdhury_2024}; \citet{D_nes_2014}.  
Although the radio and optical radiation originates in different components of galaxies, it appears that there is some underlying aspect of key processes that maintains this correlation. 
Thus further studies may help us in developing a comprehensive picture of galaxy formation and evolution. 

Redshifted $21$~cm emission serves as a key tracer of \HI gas and its dynamics in the ISM and CGM. 
Therefore, detection of a large sample of \HI galaxies, complemented by multi-wavelength data from independent surveys probing the same cosmic volume, provides a powerful framework for studying galaxy evolution. 
For example, the joint ALFALFA–SDSS optical–\HI catalogue~\citep{Durbala_2020} was constructed by cross-matching ALFALFA \HI detections within the common survey volume of SDSS. 
With sufficient redshift and environment coverage, these joint observations enable detailed investigations of the distribution and evolution of neutral gas across cosmic time and allow us to correlate galaxy properties observed across different wavelengths. 
That said, optical surveys can probe galaxy properties to great depths, for example: COSMOS-web probes the optical properties up to $z\thicksim 3-4$~\citep[e.g.,][]{Casey_2023}. 
While directly detecting the \HI $21$~cm emission at $z\gtrsim 0.3$ is challenging and requires a very long integration time, as it has a very low transition probability. 
However, \HI-stacking technique enabled measurement of average properties of galaxy population to $ z\approx 1$ \citep[e.g.,][]{Rhee2013,Bera2019,Bera_2022,Chowdhury2021,Chowdhury_2024}. 
\HI-stacking uses precise galaxy positions to detect faint \HI emission~(at the cost of losing individual source information) from distant galaxies for which signals are below the direct detection threshold of the radio instruments. 
Optical galaxy catalogues~(e.g, SDSS, COSMOS) are often used as a proxy for the location of the \HI emission to extract combined flux of such sources~(sub-threshold sources).

Ongoing surveys with the SKA precursors and upcoming surveys with the SKA are expected to extend the direct detection limit to $\ z\approx 1$. 
Sub-threshold sources in these surveys can further extend \HI studies beyond $z > 1$ through \HI stacking. 
The \textit{MeerKAT International Giga Hertz Tiered Extragalactic Exploration}~\citep[MIGHTEE-HI;][]{Maddox_2021} and the \textit{Widefield ASKAP L-Band Legacy All-Sky Blind Survey}~\citep[WALLABY;][]{Koribalski_2020}) are two such large survey projects. 
On the other hand, large multi-wavelength galaxy surveys such as COSMOS-Web~\citep{Casey_2023}, and Dark Energy Spectroscopic Instrument~\citep [DESI;][]{Hahn_2023} will cover sufficient depth and sky area to enable complementary studies of galaxy populations in optical bands.  
Given that ongoing \HI surveys are expected to detect a large number of sources with unprecedented sensitivity \citep[e.g.,][]{Duffy_2012a, Maddox_2021, Bharti_2022} and to investigate the expected outcomes (\HI and optical science) with high statistical significance from these surveys, it is essential to simulate these surveys and develop mock catalogues for these planned/ongoing surveys. 
Mock catalogues can help extract information early on, optimize data processing, and assess the capacity of observations to distinguish between different scenarios of galaxy formation. 
Various methods for constructing mock catalogues are available, with most based on halo model or semi-analytical models~\citep[e.g.,][]{Obreschkow_2009,Paul_2019,Paranjape_2021,Qin_2023} and N-body simulations~\citet{Springel}. 
Mock catalogues also help assess the effectiveness of planned survey strategies in achieving the expected scientific objectives~\citet{Maddox_2016} for example, measuring \HI mass function and optical to \HI scaling relations. 

The \HI mass function (hereafter \HIMF) provides a statistical description of how \HI gas is distributed among galaxies in the Universe. 
The second moment of \HIMF produces the \HI\ mass density, $\rho_{\rm \HI}$, whose redshift evolution is a crucial test for galaxy evolution models. 
\HIMF of the local Universe has been measured in various surveys, HIPASS~\citep{Zwaan_2005}, ALFALFA~\citep{Jones_2018} and FASHI~\citep{Ma2025}.  
Direct measurements of the \HIMF at intermediate to higher redshift are limited and biased by the small number of individual detections~\citep{Ponomareva_2023}.  
Many of the estimates make use of scaling relations and stacking and hence are unable to provide tight constraints on the parameters describing the mass function. 
The upcoming SKA pathfinders and precursor surveys have the potential to measure it across a wide range of redshifts and environments~\citep{Duffy_2012,Baker_2024}. 
One way to push the measurement below the direct detection level is to use \HI stacking~\citep[e.g.,][]{Bera2019,Bera_2022, Chowdhury2020, Chowdhury2021,Chowdhury_2024}. 
\citet{Pan2020} and \citet{WangJ_2025} used a Bayesian stacking technique to estimate \HIMF below the noise level, though their technique differs from conventional \HI-stacking and provides a robust estimate of \HIMF assuming source proxies from other spectroscopic surveys. 
Although these studies demonstrated that \HIMF can be estimated from the upcoming SKA surveys, further investigation is required to include additional information, such as inclination, size, colour, and luminosity. 
Therefore, a joint catalogue is essential for reconstructing the \HI mass function from stacking measurements for different galaxy populations~\citep{Dutta_2021} and across different redshifts.

In our current work, we generate joint mock catalogues and explore their use for predicting expected outcomes such as direct detections and stacking significance for average \HI mass in SKA precursor surveys. 
Unlike conventional approaches to creating mocks, our method is distinct in its simplicity, as it does not rely on dark matter simulations (N-body simulations) as the starting point. 
We adopt this approach because relying on the full halo distribution would necessitate abundance matching, which can be an unreliable assumption in some situations~\citep{Zentner_2014,Campbell_2018}. 
Our method is computationally efficient and offers quick ways to explore detection prospects and galaxy evolution. 
Our mock catalogues account for most of the aspects of actual observations except clustering. 
This does not affect our predictions as long as we target larger surveys: MIGHTEE-HI, LADUMA and WALLABY. 
We also assign optical colours and luminosities to the sources in the mock catalogues using a simple, novel, and observation-inspired approach. 
Incorporating optical properties significantly enhances their potential for investigating galaxy evolution and connecting multi-wavelength properties of galaxies~\citep[e.g.,][]{Dutta_2020, Dutta_2021, Li_2022, Lu_2024}.
Mock observables in our catalogues: the number of direct detections and average stacked \HI mass can be used to estimate the \HIMF accurately using Bayesian analysis. 
We demonstrate that combining stacking measurements with direct detection puts stronger constraints on the \HI mass function, especially at higher redshifts.
Although we adopt standard techniques for stacking and constraining the \HIMF parameters, this work explore three new aspects: (i) for the first time, we introduce a novel approach (see \Sref{ssec:alfalfa_mock}) for constructing joint optical–\HI catalogues based on local observations, by estimating pixel-wise \HIMFs across the colour–magnitude plane. The resulting joint catalogue provides a framework for evaluating the potential of upcoming and ongoing surveys to measure the optical-\HI properties at the expected significance. (ii) a quantitative stacking analysis across colour-magnitude bins and (iii) for the first time, we also investigate the joint constraints on \HIMF recovery by combining direct detections and \HI-stacking in upcoming and ongoing surveys.

This work is organised as follows. 
\Sref{sec:A100-SDSS_cat} briefly reviews the observational data (A100-SDSS) used in this work as template. 
In~\Sref{sec:mock_cat}, we discuss our method to create a joint catalogue and validate our approach with local observational data. 
We also present a method to create a joint catalogue for SKA precursors in the same section. 
Results are presented in \Sref{sec:result}. 
We conclude and summarise our work in \Sref{sec:conclusions}. 
Throughout this work, we assume a flat $\Lambda$CDM cosmology described by, $(\Omega_m, \Omega_{\Lambda}, h) = (0.3, 0.7, 0.7)$ to calculate redshifts, distances, and absolute magnitudes.

%%%%%%%%%%%%%%%%%%%%%%%%%%%%%%%%%%%%%%%%%%%%%%%%%%%%%%%%%%%%%%%%%%%%%%%%

%%%%%%%%%%%%%%%%%%%%%%%%%%%%%%%%%%%%
\begin{figure*}
    \centering
    \includegraphics[width=0.9\linewidth]{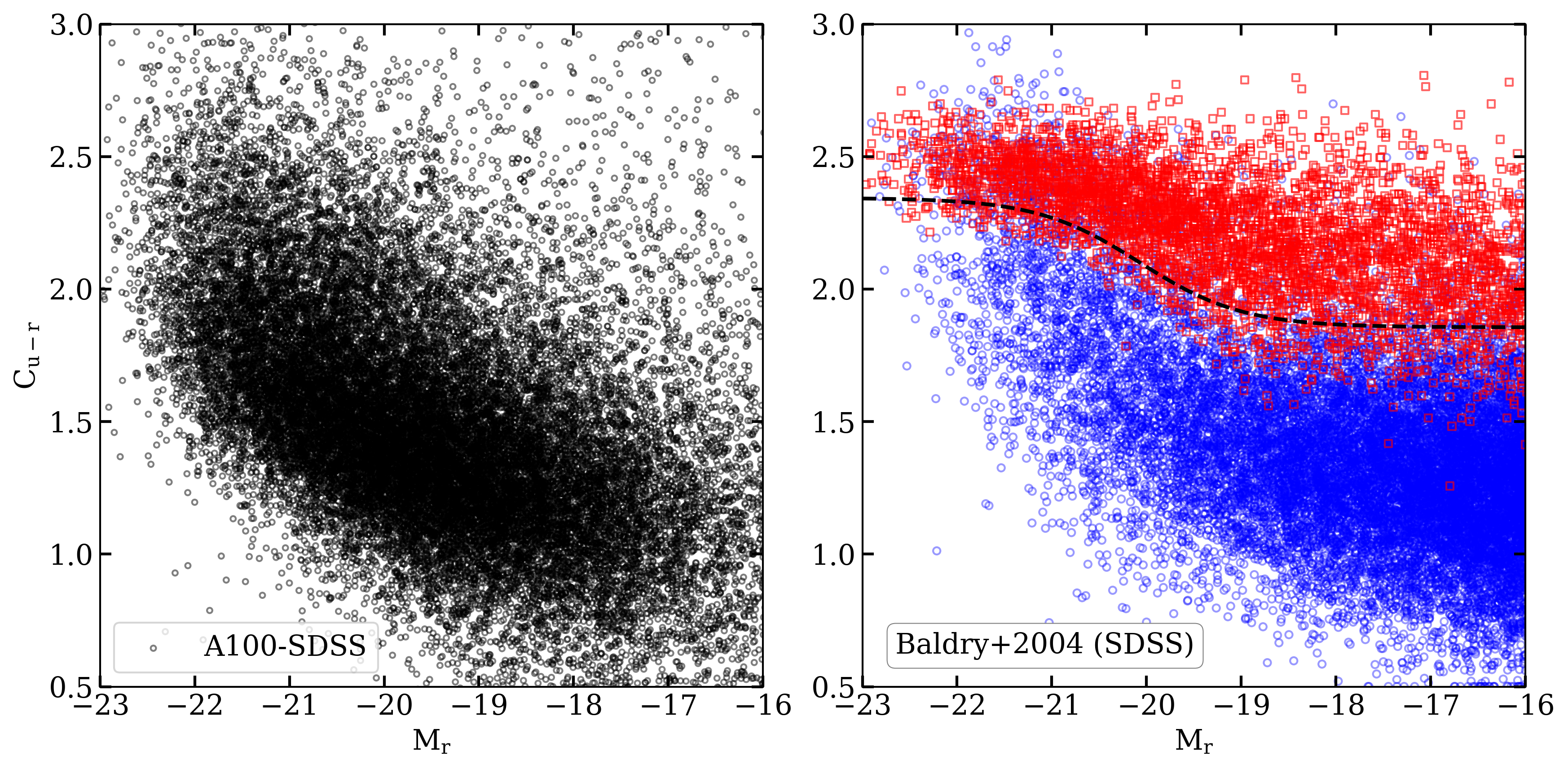}
    \caption{Distribution of galaxies in the colour-magnitude plane: In the \emph{left panel}, we show sources in A100–SDSS catalogue, while the \emph{right panel} presents a volume-limited sample that displays the expected bimodality in colour and magnitude which we reproduced following the method discussed in~\citet{Baldry}. A black dashed line marks the optimal divider that separates two populations of galaxies. The number of sources differs in the two panels as they represent flux-limited and volume-limited samples, respectively.}
    \label{fig:cmplane}
\end{figure*}
%%%%%%%%%%%%%%%%%%%%%%%%%%%%%%%%%%%%%
 
%%%%%%%%%%%%%%%%%%%%%%%%%%%%%%%%%%%%%%%%%%%%%%%%%%%%%%%%%%%%%%%%%%%%%%%%

\section{DATA: ALFALFA-SDSS joint catalogue}
\label{sec:A100-SDSS_cat}

This section briefly describes observational inputs used to generate a mock catalogues for the ALFALFA-SDSS survey, which we subsequently do in later sections. 
In the ALFALFA survey, $31,500$ \HI sources have been detected over the $\thicksim7000$ deg$^2$ survey area on the sky, up to the redshift of $z\thicksim0.06$ \citep{Giovanelli_2005,Haynes}. 
The ALFALFA-HI survey data releases: $\alpha.40$, $\alpha.70$ and $\alpha.100$ helped in studying different aspects of galaxies and their evolution. 
These studies involve \HI mass function (\HIMF), \HI velocity width function, \HI scaling relation between gas fraction, \HI mass and \HI selected galaxy correlation function etc.~\citep[e.g.,][]{Haynes_b, Jone_2016, Janowiecki_2019, Haynes_2011, Papastergis_2013}. 
Besides focusing on \HI alone, ALFALFA also enables us to study and compare the properties of galaxies at other wavelengths. 
This can be achieved by identifying counterparts of \HI detections of ALFALFA in the common volume of other surveys, e.g., the Sloan Digital Sky Survey~(SDSS)~\citet{Abazajian_2003}. 
Using the $\alpha.40$ sample with optical counterparts in SDSS,~\citet{Dutta_2020,Dutta_2020a} established a correlation between \HI distribution and optical properties, and \citet{Huang_2012} estimated scaling relations between gas fraction and optical and UV colours.

We work with optical-\HI catalogue of $\thicksim 30,000$ galaxies obtained from the $100\%$ complete ALFALFA catalogue (hereafter, A100-SDSS catalogue)~\citet{Durbala_2020}. 
The A100-SDSS catalogue lists \HI properties for local galaxies together with their SDSS optical counterpart object IDs. 
We refer the reader to $\S 4.1$ of \citet{Haynes_2011} for details on assigning optical counterparts to ALFALFA detections. 
\textcolor{blue}{Tanya et al.~2026 (in prep.)} used optical counterpart object IDs to obtain the corresponding SDSS photometric properties: \emph{ugriz} values (model magnitudes and their uncertainties). 
Model magnitudes are corrected for Milky Way extinction~\citep{Schlegel_1998}. 
Rest-frame magnitudes are obtained using \textsc{KCORRECT}\footnote{\url{https://kcorrect.readthedocs.io/en/stable/}}~\citep{Blanton_2007}. 
Sources for which spectroscopic redshifts are not available, \HI redshifts are used for \textsc{KCORRECT}. 
Obtained final catalogue (A100-SDSS) is from the coverage of $\thicksim 6900$ deg$^2$ with the comoving volume of $\thicksim 6.65\times 10^{6}$ Mpc$^3$ and contains the following quantities: (i) A unique AGC ID, (ii) RA and Dec, (iii) heliocentric velocity, (iv) \HI linewidth $W_{50}$, (v) \HI Integrated flux $S_{21}$, (vi) \HI mass $M_{\rm \HI}$ (vii) distance $D$, (viii) SNR, (ix) Absolute magnitudes~($M_u, M_g, M_r, M_i, M_z$) and (x) Code (1,2). 
The Code 1 and Code 2 objects correspond to sources with $\rm SNR > 6.5$ and $\rm SNR < 6.5$, respectively. 
In current work, we work mostly with code 1 sources (signal-to-noise ratio $ > 5.0$) as the reliability of these sources is $ > 90\%$ \citep{Saintonge_2007}. 
This corresponds to a sample containing $\thicksim 94\%$ of all the total sources in the chosen colour-magnitude plane. 
\emph{Left panel} of \Fref{fig:cmplane} shows the optical properties of the A100-SDSS catalogue, obtained from $100\%$ complete ALFALFA catalogue, while the right panel presents a volume-limited sample of SDSS optical galaxies reproduced using the methods given in \citealt{Baldry}~(see next section for details).

%%%%%%%%%%%%%%%%%%%%%%%%%%%%%%%%%%%%%%%%%%%%%%%%%%%%%%%%%%%%%%%%%%%%%%%%
%%%%%%%%%%%%%%%%%%%%%%%%%%%%%%%%%%%%%%%%%%%%%%%%%%%%%%%%%%%%%%%%%%%%%%%%

\section{Joint Mock Catalogues}
\label{sec:mock_cat}

In this section, we describe the method to correlate \HI mass to optical properties of galaxies, such as r-band luminosity and their colours. 
In order to make predictions for SKA precursors, we first simulate mock data that mimic the MIGHTEE-HI and WALLABY surveys for the observing bands of MeerKAT and ASKAP arrays.  
We begin by simulating the ALFALFA \HI survey and validating it with observational data. 
We also simulate joint optical-\HI surveys with the sensitivity of ALFALFA to reconstruct A100-SDSS mock catalogue. 
This is done to check the robustness of the reproduced correlation between \HI mass and optical properties. 
We estimate \HIMF for the joint mock catalogue and compare it with that of data (A100-SDSS). 
We provide details of these steps in the following subsections.

\subsection{ALFALFA: joint mock catalogue}
\label{ssec:alfalfa_mock}

\begin{figure}
\centering
\begin{minipage}[b]{3.5in}  % Match the width of the figure
\centering
    \includegraphics[width=\textwidth]{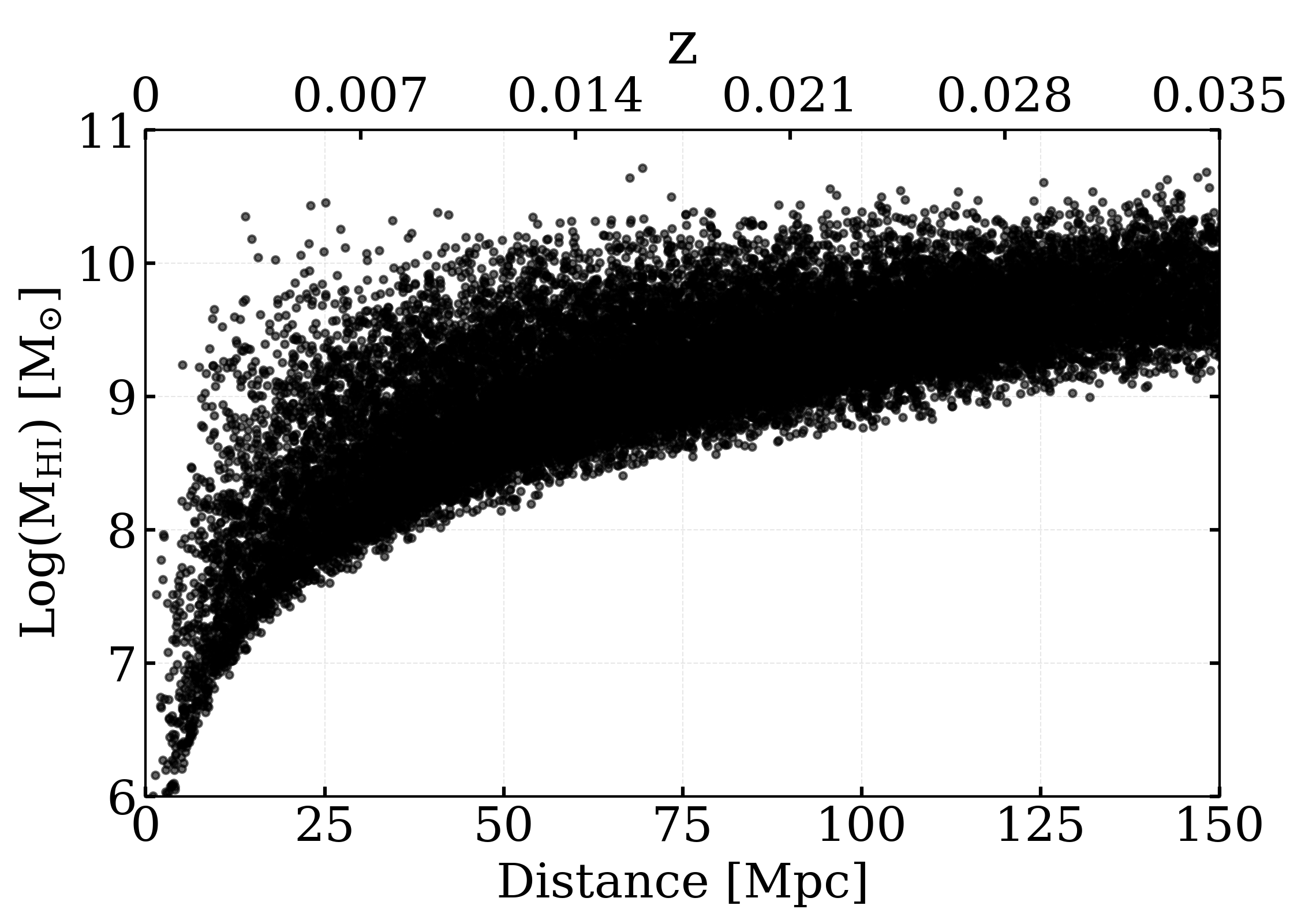}
    \caption{Characteristics of sources in the ALFALFA-\HI mock catalogue. The redshift and \HI mass are plotted for 24000 sources. This illustration is for validation purposes with the number counts within the volume simulated in \citep{Giovanelli_2005}.}
    \label{fig:RS02_prediction}
\end{minipage}
\end{figure}

To simulate the ALFALFA survey in both \HI-only and joint optical-\HI modes, we randomly distribute galaxies within an observing cone defined by the telescope’s field of view (FoV). 
To generate the redshift of the source, we use the random relative volume corresponding to the comoving distance at redshift $z$. 
The \HI mass function of the local universe is normalised and taken as the probability distribution function to assign the \HI masses to the sources within the cone. 
We assume that the \HI mass function follows the well-known Schechter form~\citep{Schechter}, which is given as
\begin{equation}
    \phi(M_{\rm \HI}) = ln(10)\,\phi^* \bigg(\frac{M_{\rm \HI}}{M_{\rm \HI}^*}\bigg)^{\alpha + 1} \exp{ \bigg(\frac{-M_{\rm \HI}}{M_{\rm \HI}^*}\bigg)}
    \label{eq:himf}
\end{equation}
where $\alpha$ is the lower mass end slope, $M_{\rm \HI}^*$ is knee mass in unit of $\rm M_{\odot}$ and $\phi^*$ (per $\rm Mpc^{3}$) is the normalization. 
Once we have redshifts and \HI masses of the galaxies in the cone, the next step is to assign the other observable quantities such as linewidth, inclination and \HI disk size. 
For \HI linewidth at $20\%$ of peak flux, $W_{20}$, we use the Baryonic Tully-Fisher relation \citep[BTFR;][]{McGaugh_2000}. 
The inclination effect is accounted for in line of sight such that the $\cos(I)$ is distributed uniformly in the range $[0,1]$. 
For the disk size we use \HI size-mass $(D_{\rm \HI}-M_{\rm \HI})$ relation from~\citet{WangJ_2025}. 
First, we perform \HI only mock observation for ALFALFA sensitivity to validate our predictions. 
We compute the signal-to-noise ratio threshold according to Eq. (4) in \citet{Giovanelli_2005}. 
For a source of \HI mass $M_{\rm \HI}$ at distance $D$ with linewidth $W_{20}$, the SNR threshold of $6$ can be given as follows:
\begin{equation}
    12.3\,f_{\beta}\,\sqrt{t_s}\,\bigg(\frac{M_{\rm \HI}}{10^6 \,\rm M_{\odot}} \bigg) D^{-2}_{\rm Mpc} \bigg(\frac{W_{20}}{200 \,\rm km/s} \bigg)^{\gamma} > 6 ,
    \label{eq:ALFA_sensitivity}
\end{equation}
where $\gamma = -1/2$ for $W_{20} < 200$ and $\gamma = -1$ for  $W_{20} \ge 200$; $t_s$ is integration time of $30$ sec per pixel solid angle; and $f_{\beta} \leq 1$ determine the fraction of source flux detected by the Arecibo's beam. 
\citet{Giovanelli_2005} simulated the ALFALFA \HI surveys by assuming two \HIMFs: RS02 \citet{Rosenberg_2002} and Z97 \citet{Zwaan_1997}. 
We take RS02 \HIMF to validate our prediction for number counts detected in volume within $\mathrm{distance} \le 150 $~Mpc. 
The expected number of detections with RS02 \HIMF is shown in~\Fref{fig:RS02_prediction}. 
Their expected number of detection is $22,200$ with RS02 \HIMF, which is slightly lower than our prediction $\thicksim 24000$. 
Our estimates are conservative, and unlike them, we do not consider clustering in space densities.

%%%%%%%%%%%%%%%%%%%%%%%%%%%%%%%
%%%%%%%%%%%%%%%%%%%%%%%%%%%%%%%

Next, we simulate the joint optical–\HI survey by assigning each source a colour–magnitude value  corresponding to its \HI mass. 
As discussed, galaxies have a bivariate distribution in colours, and the distribution of colour is also a function of the magnitude. 
\citet{Baldry} quantified the colour and magnitude distribution using the SDSS observations and derived the relations between (u-r) colours and r-band magnitude. 
They defined the colour-magnitude distribution as the sum of two distributions: red and blue distributions $\phi_{\rm comb}=\phi_r + \phi_b$ such that $\phi_{\rm comb} dM_r dC_{ur}$ is the number of galaxies between r-band magnitude $M_r$ and $M_r+dM_r$ and between colour $C_{ur}$ and $C_{ur}+dC_{ur}$. 
The parameterization given by them is as follows: 
\begin{equation}
	\phi_{\rm comb}(M_r,C_{ur}) = \phi(M_r)\, \mathcal{G}(C_{ur},\mu(M_r),\sigma(M_r))
\end{equation}
where $\mathcal{G}$ is a Gaussian distribution with mean $\mu$ and scatter $\sigma$; both are constrained to be contiguous function of $M_r$.
\begin{equation}
	\mathcal{T}(M_r) = p_0 + p_1(M_r+20)+q_0\tanh\left(\frac{M_r-q_1}{q_2}\right)
\end{equation}
where $p_1,p_2$ and $q_1,q_2$ are fitted parameters for the two populations of galaxies. 
They fitted a Schechter form \citep{Schechter} of luminosity function in term of magnitude, given as,
\begin{equation}
  	\phi(M_r) = c\phi^*e^{-c(\alpha+1)(M_r-M^*)}e^{-e^{-c(M_r-M^*)}}
\end{equation}
where $c=0.4\ln(10)$; $M^*$~and~$\phi^*$ are characteristic number density and magnitude and $\alpha$ is faint-end slope. 
We take these parametrised relations and fitted values of  $p_1,p_2$ and $q_1,q_2$ (see table 1 of \citet{Baldry} for details) and generated a volume-limited sample of galaxies. 
The bimodal colour $(u-r)$ magnitude $M_r$ distribution for this mock catalogue is shown in the right panel of~\Fref{fig:cmplane}. 
This distribution represents the sample of galaxies observed in a volume (within $z\leq 0.05$) of the SDSS survey. 
Note that the \HI selected galaxies in the A100-SDSS catalogue (\emph{left-panel}) are dominantly blue galaxies. 
The generated SDSS galaxy distribution (\emph{right panel}) in the colour $C_{u-r} = M_u-M_r$ and r-band magnitude $M_r$ plane is bimodal, which constitutes: (1) The blue galaxies (mostly star-forming main sequence galaxies) and (2) Red galaxies (passive \HI deficit or quenched galaxies). 
The optimal colour divider, indicated by the \emph{black-dashed} line, separates the galaxies into two distinct populations. 
The colour distribution of galaxies in the A100-SDSS catalogue shows a bias due to \HI selection when compared to the optical distribution shown in the \emph{right panel}. 
This is due to the non-detection of red galaxies in ALFALFA-\HI survey. 
Since the colour–magnitude (CM-plane) sample ( in \emph{right panel}) is constructed from the optical survey alone, using it in joint optical-\HI surveys results in a biased \HI selection, and can not be used to correlate its optical properties to \HI properties of A100-SDSS data. 
Generated SDSS optical galaxy sample (volume limited) in \emph{right-panel} is shown for comparison purposes to illustrate bias in ALFALFA \HI detection (\emph{left-panel}).
Thus we need to work with the joint catalogue and not the optical data alone for construction of joint mock catalogue.

To construct the CM plane for the joint optical-\HI survey, we take the A100-SDSS catalogue as a template. 
To generate the CM plane for the joint \HI optical survey, we follow these steps,

%%%%%%%%%%%%%%%%%%%%%%%%CM planes%%%%%%%%%%%%%%%%%%%%%
\begin{figure*}
    \centering
    \includegraphics[width=0.9\linewidth]{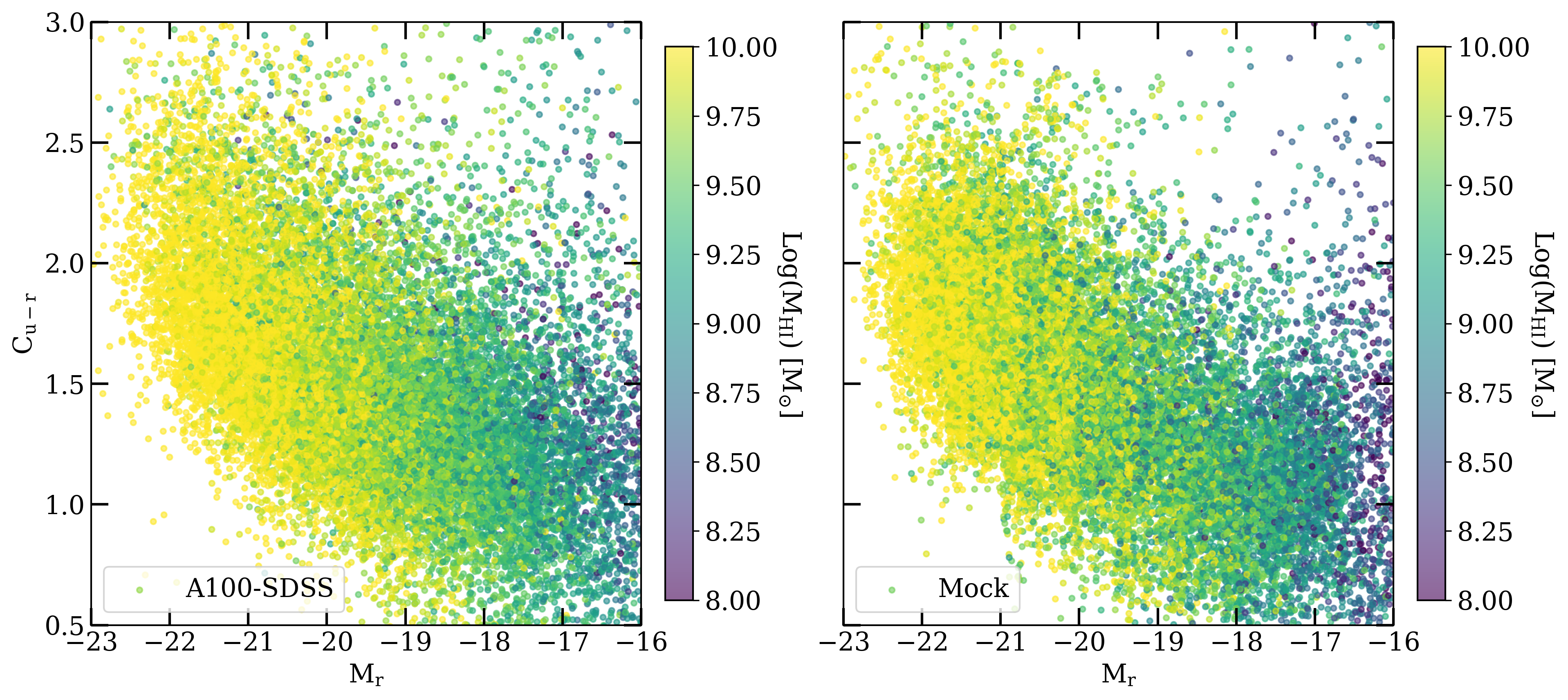}
    \caption{ Comparison of mock catalogue with the input A100-SDSS catalogue~(at $z\thicksim 0$): \emph{Left panel} shows the optical properties of galaxies in A100-SDSS catalogue, where r-band magnitude is displayed on the \emph{x-axis} and the (u-r) colour on the \emph{y-axis}. Sources are colour-coded according to their \HI mass. \emph{Right panel} illustrates the corresponding properties for galaxies in our mock catalogue, obtained from the ALFALFA joint survey described in~\Sref{ssec:alfalfa_mock}.}
    \label{fig:CM_plane}
\end{figure*}
%%%%%%%%%%%%%%%%%%%%%%%%%%%%%%%%%%%%%%%%%%%%%%%%%%%%%%%

\begin{itemize}
    \item As a first step, we estimate the r-band luminosity function from the A100-SDSS catalogue. 
    Since the data has a joint \HI and optical(SDSS) selection, we use the \HI flux and r-band flux to calculate the maximum volume in which a galaxy is detectable by the SDSS and ALFALFA survey, given its optical and \HI flux. 
    This maximum volume for the galaxies can be calculated as, $V_{\rm max} = min(V^{\rm Opt}_{\rm max},V^{\rm \HI}_{\rm max})$. 
    The joint luminosity function (LF) can be estimated using $1/V_{\rm max}$ method \citep{Schmidt}. 
    We fit the luminosity function using the Schechter function \citep{Schechter}, and our best-fit parameters are consistent within $2\sigma$ uncertainty with what was estimated by \citet{Baldry}. 
    We refer reader to~\Aref{ssec:fit_lum_clr} and \emph{left panel} of~\Fref{fig:A100SDSS_LF_CM} for details of $1/V_{\rm max}$ method.
    
    \item Next, we divide the sample into $10$ bins of width $\Delta M_r = 0.7$ in absolute magnitude, and fit a skewed Gaussian to the $V/V_{\text{max}}$ corrected number counts of $(u - r)$ colours in each bin. 
    The motivation for using a skewed Gaussian is that the ALFALFA galaxy sample is \HI selected, which causes the typically bimodal Gaussian colour distribution to appear as a single-skewed distribution. 
    For further details, the reader is referred to the right panel of \Fref{fig:A100SDSS_LF_CM} in~\Aref{ssec:fit_lum_clr}, which shows the different fits to the (u–r) colour number counts.
    
    \item The best-fit parameters of the skewed Gaussian are used to draw random $(u - r)$ colours within each magnitude bin, populating the CM-plane pixel by pixel.
\end{itemize}
 
Once we have generated colour and magnitude for each galaxy in our mock catalogue, the next step is to assign \HI masses to them. 
One can assign \HI mass to each galaxy using the derived CM and \HI mass relations. 
We use the A100-SDSS catalogue to establish relations between the \HI mass and colour-magnitude. 
To establish these relations and assign \HI mass to galaxies in CM plane, we follow these steps:
 \begin{itemize}
     \item In a $ 4\times 4$ grid within the CM plane, we compute \HI mass functions for each cell. 
     We fit Schechter parameters $(\alpha, M^*_{\rm \HI},\phi^*)$ for the \HI mass function within each cell.

     \item We assign these parameter values to the centre of each cell. 
     To achieve finer variation of Schechter parameters in the  $(C_{ur} - M_r)$ plane, we use 2D linear interpolation.

     \item The \HI masses are then assigned to galaxies within each fine pixel $(\Delta C_{ur}, \Delta M_r)$ based on the corresponding \HIMF.

     \item On the fine grid of the CM plane, this approach allowed us to establish the relations between the \HI mass of the galaxies and their optical properties. 
     Once we have these relations, they can be applied to any joint optical-\HI mock survey after validation of the scheme.
 \end{itemize}

Using the $\alpha.40$ sample, \citet{Saili} estimated the conditional \HI mass function (\HIMF) by dividing galaxies into different populations based on colour and magnitude cuts. In contrast, using the larger ALFALFA $100\%$ catalogue, we compute the \HIMFs on a fine, pixel-by-pixel basis across the colour–magnitude space, rather than relying on discrete colour–magnitude classifications. This approach can connect the \HI properties to the optical properties of galaxies within the common volume of optical and \HI surveys (e.g., ALFALFA and SDSS in current work) and enable us to study their evolution.
In flux-limited surveys, sources with integrated flux $S_{\rm 21}$ are detected over a certain range of line widths $ W_{50}$ as detection threshold scales as $\mathrm{S/N}\propto S_{\rm 21}/\sqrt{W_{50}}$. 
The completeness of the ALFALFA survey in the ${S_{\rm 21}-W_{50}}$ plane was quantified by \citet{Haynes_2011} using the galaxy sample itself. 
This completeness is described by a smooth function, $C(S_{\rm 21}W_{50})$, which characterises the survey sensitivity as a function of integrated flux and linewidth. 
As the A100–SDSS catalogue is flux-limited, we apply the ALFALFA completeness cut to estimate the \HIMF. 
We use the following completeness cut derived by \citet{Ma2025},
\begin{equation}
    \log S_{21} = \begin{cases}
        0.5\log W_{50}-1.162 & \text{: } \log W_{50} < 2.476 \\
        \log W_{50}-2.40 & \text{: } \log W_{50} \geq 2.476
    \end{cases}
\end{equation}
We estimated the global \HIMF for the A100-SDSS catalogue without any optical cuts and validated it with that of \citet{Ma2025}. 
Our \HIMF estimate is consistent with their result in the declination range $0^{\circ} < \text{Dec} < 30^{\circ}$, covering a sky area of $ 5649\,\text{deg}^2$.

It is important to note that we chose to employ the traditional $1/V_{\rm max}$ \citep{Schmidt} method to estimate \HIMF for each pixel. 
However, the $1/V_{\rm max}$ method is not robust against the fluctuations due to the large-scale structure compared to commonly used methods: $1/V_{\rm eff}$ method \citep{Martin_2010} and the two-dimensional stepwise maximum likelihood(2DSWML) method \citep[e.g.,][]{Zwaan_2005,Martin_2010,Efstathiou_1988,Jones_2018}. 
Since, we do not account for clustering in our models for creating a mock catalogue, so with a very large volume in our simulation, the $1/V_{\rm max}$ method can suppress the effect of large-scale structure in the \HIMF estimation, and we do not expect it to affect our results in a significant manner. 
The other issue of $1/V_{\rm eff}$ and the 2DSWML method is their high sensitivity towards the exact completeness cut. 
The sensitivity of the completeness cut becomes much severe with the low number counts of \HI sources within the colour-magnitude pixel. 
This leads to larger uncertainties and biases in \HIMF~ estimation. 
We therefore adopt the $1/V_{\max}$ estimator to compute the pixel-wise \HIMFs.

\begin{figure}
\centering
\includegraphics[width=0.91\linewidth]{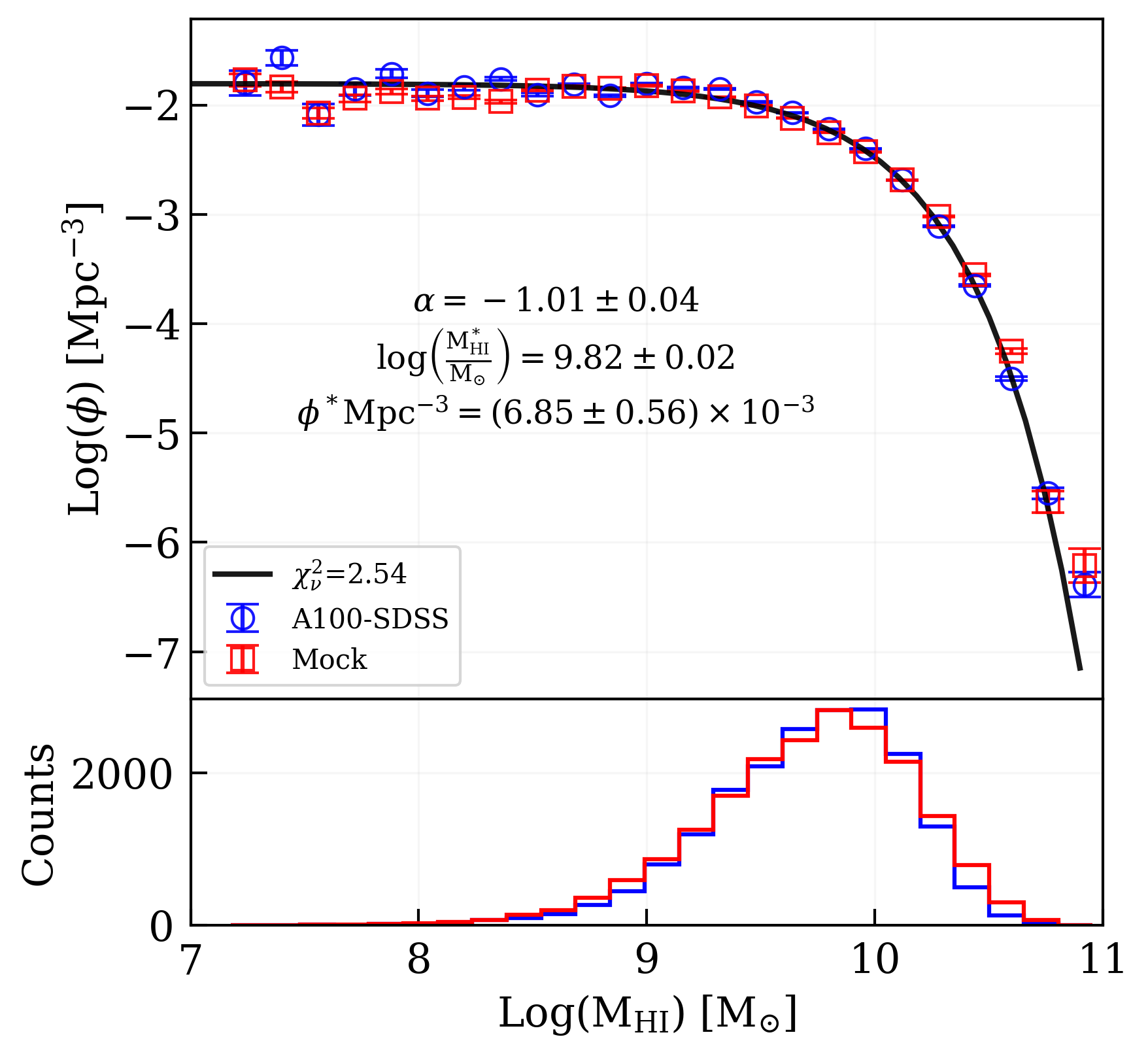}
\caption{The comparison of the mock \HIMF in the A100-SDSS catalogue. \emph{Red open squares} show the \HIMF estimated from our mock catalogue for the joint survey. \emph{Blue open circles} represent the \HIMF estimated from the A100-SDSS catalogue. Both estimates are obtained using the $1/V_{\rm max}$ method. Displayed Schechter fitting parameters correspond to the mock catalogue. \emph{Bottom panel} presents the distribution of \HI masses for both the mock catalogue and the observational catalogue.} 
\label{fig:HIMF_alfa}
\end{figure}

To evaluate the robustness of the relation between \HI mass and colour-magnitude, which we set up above, we simulate joint optical–\HI surveys using the ALFALFA sensitivity. 
For this, we assign \HI masses to galaxies according to the \HIMF corresponding to their location in the colour-magnitude plane. 
We adopt the ALFALFA flux sensitivity (Eq.~\ref{eq:ALFA_sensitivity}) with a $30$ sec integration time per beam solid angle. 
The joint ALFALFA-SDSS mock survey result is presented in the right panel ~\Fref{fig:CM_plane}. 
Comparison of mock \HI mass function with that of A100-SDSS data is shown in ~\Fref{fig:HIMF_alfa}. 
Here, we compare the \HI mass function derived from the mock catalogue (red open squares) with that of the A100–SDSS data (blue open circles). 
The mock \HI mass function and color–magnitude distribution are consistent with the observed one, indicating that our approach effectively captures the underlying relationship between \HI mass and optical properties. 
Although a limited number of sources in some pixels introduce large uncertainties in the local \HI optical relations, the impact on the global correlation is negligible. 
Given the large survey volume of our mock catalogues, we do not expect this to affect our predictions. 
Future radio and optical surveys will observe larger samples and populate the colour–magnitude plane more uniformly; therefore, with the available A100–SDSS data, the present approach illustrates the current state of the art. 
Therefore, our method can be used for generating a joint \HI optical mock catalogue with sensitivity limits of any telescope. 

%%%%%%%%%%%%%%%%%%%%%%%%%%%%%%%%%%%%%%%%%%%%%%%%%%%%%%%%%%%%%%%%%%%%%%%%
\subsection{SKA precursors: joint mock catalogue}
\label{ssec:ska_prec}

In this section, we briefly describe our models for generating the mock catalogue when accounting for the optimal sensitivity of the radio arrays. 
To compute the sensitivity, we estimate the effective number of baselines $N_B$ of the radio array that contribute at the required resolution. 
We generate the U-V plane coverage taking declination of the source and latitude of the array, including variation in projected baselines due to Earth's diurnal rotation.  
We keep only baselines that just resolve the source, discarding longer ones to avoid over-resolving the source and losing signal. 
For a given source size $\theta_{\rm size}$, effective baselines (with baseline length $B = \sqrt{u^2+v^2}$) are those for which $B\leq B_{\rm crit}$. 
The critical baseline $B_{\rm crit}$ is the baseline at which source size matches the angular resolution of the array:
\begin{equation}
    \theta_{\rm size} = 1.22\frac{\lambda_{\rm obs}}{B_{\rm crit}}
\end{equation}
Where $\lambda_{\rm obs}$ is the observed wavelength of the emission. 
The effective number of such baselines is denoted by $N_B$, and they are used to estimate the sensitivity of the array for the given source size. 
The expected thermal noise, i.e. \textit{rms} noise, in the receiver system can be estimated as (in Jy/synthesised beam),
\begin{equation}
    \sigma_{rms}=\frac{(T_{\rm sys}/G)}{\sqrt{2N_{B}\,\Delta t\,\Delta \nu}}
    \label{eq:noise}
\end{equation}
where $T_{sys}$ is system temperature, $\Delta t$ is the integration time, $G=A_e/2k$ is antenna gain, $A_e$ is the effective area of each antenna and $k=1390~\mathrm{Jy.m^2/K}$ is the Boltzman constant. 
For sources with a small angular extent: either smaller galaxies or distant galaxies, where the subtended angle is smaller than the resolution of the array, all baselines are used. 
$\Delta \nu$ is the frequency width over which the flux of the source is distributed. 
A larger frequency width optimises detection sensitivity by capturing the entire spectral line over a bandwidth. 
We account for the random inclination effect to the line of sight such that $\cos(I)$ is distributed uniformly in $[0,1]$. The linewidth $W_{20}$ for source having a line of sight orientation $I$ can be give as follows:
\begin{equation}
    W_{20}=2\sqrt{V_c^2\sin^2{I}+\sigma_{v}^2}
\end{equation}
where $V_c$ is the rotation velocity depend on \HI mass via Baryonic Tully-Fisher Relation~\citep[BTFR;][]{McGaugh_2000} and $\sigma_{v}=10$ km/s is dispersion. 
Note that throughout our simulations, we adopt the linewidth measured at $20\%$ of the peak flux, rather than at $50\%$ as used in the A100–SDSS catalogue. The expected signal (flux density $S_{v}$) of a galaxy at redshift $z$, with HI mass $M_{\rm \HI}$ and velocity width $W_{20}$ can be computed using:
\begin{equation}
    S_{v} \approx \frac{\beta(\theta)}{(2.36\times\,10^5\,W_{20})}\frac{M_{\rm \HI}\,(1+z)}{D^2_L}
    \label{eq:HIsignal}
\end{equation}
where $D_L$ is the luminosity distance of the galaxy and $\beta(\theta)$ is the primary beam pattern of the antenna. 
We refer reader to~\citet{Bharti_2022} for more details about sensitivity parameters, signal, primary beam patterns and other parameters for MeerKAT and ASKAP. 
In our simulated survey for SKA precursors, we use the \HI mass and optical colour-magnitude relations. 
The resulting mock catalogues include both \HI and optical properties, enabling further analysis such as direct detection above a specified signal-to-noise ratio (SNR), optically selected \HI stacking, and assessing the sensitivity of the \HIMF against these observables. 
Observables like the number of direct detections and the stacked average \HI mass for a specific colour-magnitude pixel are valuable for constraining the \HI mass function within that pixel.

% Stacking
%%%%%%%%%%%%%%%%%%%%%%%%%%%%%%%%%%%%%%%%%%%%%%%%%%%%%%%%%%%%%%%%%%%%%%%%
We model the \HI spectral lines in order to estimate significance of \HI stacking detection. 
Stacking is the co-addition of \HI spectra for galaxies at known redshifts, where the combined SNR for the $21$~cm line increases by $\sqrt{N}$ in an ideal case. 
It is applied to sources that are below the detection limit to estimate the average \HI emission of the source population below the detection threshold. 
The simulated line profile (\HI spectral line) for the $i^{th}$ source can be given as the sum of the model plus Gaussian random noise: $S_i = S^m_i + \mathcal{N}(0,\,\sigma^{2}_{\,i,rms})$. 
We model the \HI line profile $S^m$ as a busy function \citep{Westmeier} which consists of five free parameters: the width of the profile $W_{20}$; the total amplitude scaling factor $a$; and three additional parameters $b_1$,$b_2$ and $c$. 
The parameters $b_1$,$b_2$~control the steepness of the two error functions that make the line profile flanks and horns and $c$ controls the central trough's amplitude. 
We model the \HI signal as a function of channel velocity $v$ as follows:
\begin{eqnarray}
S^{m}(v) &=& \displaystyle\frac{a{}}{4} \times (\mathrm{erf}[b_{1} \lbrace W_{20} + v \rbrace ] + 1) \nonumber \\
&& \times (\mathrm{erf}[b_{1} \lbrace W_{20} - v \rbrace ] + 1) \times \left( c{} \, v ^{2} + 1 \right) \!
\end{eqnarray}
We chose to work with the simple model of a box-car profile for low SNR sources by setting  $b_1=b_2 = 0.3$ and $c = 0$. 
The scale factor is computed using the \HI mass of the source. 
The width of the profile is taken as line width $W_{20}$ in mock catalogue. 
A given number of \HI spectra is stacked using galaxy weight $w$ equal to the inverse of the square of the spectrum rms noise. 
We add a velocity jitter of $\delta v = 10~\mathrm{km/s}$ to the velocity grid of each spectral line to account for redshift uncertainty. 
Before stacking, each spectrum is interpolated onto a common velocity grid so that all spectra are sampled at the same velocity resolution, which we take $dv \approx 10$~km/s. 
The expected average signal with weighted mean stacking is as follows,
\begin{equation}
    \langle S \rangle = \frac{\sum_i^N S_iw_i}{\sum_i^Nw_i}
\end{equation}
where $w_i = 1/\sigma_{i\, ,rms}^2$ is the weight for a given galaxy and $\sigma_{i\,,rms}$ is Gaussian random noise for  $i^{th}$ galaxy. For the \HI stacking, the COSMOS optical flux cut in i-band ($i < 26.7$) is applied to the joint catalogue. Within the range $-23 < M_r < -16$, the COSMOS's i-band and r-band absolute magnitudes follow a tight linear relation, so applying a cut in i-band or in r-band is effectively equivalent for our sample.

%%%%%%%%%%%%%%%%%%%%%%%%%%%%%%%%%%%%%%%%%%%%%%%%%%%%%%%%%%%%%%%%%%%%%%%%
\section{Model and likelihood function}

 In this section, we derive the priors and posterior probabilities for \HIMF parameters. 
 In \citet{Bharti_2022}, we performed the sensitivity analysis for \HIMF parameters against mock observables, working with the number of direct detections. 
 The sensitivity of the parameters showed that, in principle, the \HI mass function can be reconstructed using mock observables. 
 In this work, we assume a Schechter form~\Eref{eq:himf} of \HIMF and use the the number of direct detections and the average stacked \HI mass as observables in our mock surveys. 
 According to Bayes' theorem, the posterior probability of a model given data is proportional to the likelihood, which quantifies how probable the data are under that model. We define likelihood for number counts of direct detection above $5\sigma$ detection threshold with the assumption that number counts are independent in different mass bins. 
\begin{equation}
    \mathcal{L} \propto \prod_{i=\text{bins}} P_i (N_i|\mu_i)
\end{equation}
where $P_i (N_i|\mu_i)$ is the probability of getting $N_i $ observed number counts in $i^{th}$ mass bins given the model for the direct detection $\mu_i$. 
The number counts in a given mass bin follow the Poisson distribution, and their likelihood can be given as,
\begin{equation}
    P_i (N_i|\mu_i) = \frac{\mu_i^{N_i}\, e^{-\mu_i}}{N_i!}
\end{equation}
The denominator term in the log-likelihood can be ignored for the purpose of finding the maximum  as it does not depend on the model parameters. 
The log-likelihood for the number of direct detections can be given as follows,
\begin{equation}
     \ln \mathcal{L} = \sum_i^n(N_i\ln\mu_i-\mu_i)
\end{equation}
The log likelihood for the mean \HI mass is defined as the sum over the different observing fields of the difference between model $\bf{M}$ and the mean stacked \HI mass $\langle M_{\rm \HI}\rangle$.
\begin{equation}
     \ln \mathcal{L} = \sum_i^{F_N} \bigg(\frac{\langle M_{\rm \HI}\rangle- \langle \boldsymbol{M}(\alpha,M^*_{\rm \HI},\phi^*)\rangle}{\sigma_{\langle M_{\rm \HI}\rangle}}\bigg)^2
\end{equation}
where $\sigma_{\langle M_{\rm \HI}\rangle}$ is the uncertainty in the mean \HI mass and $F_N$ is number of observing fields used for \HI stacking. 

In our mock catalogue, the expected number of direct detections ($\rm SNR > 5.0$) in a given mass bin can also be computed by the following triple integral, where we assume the Schechter \HIMF model \Eref{eq:himf},
\begin{eqnarray}
    \mu_i &=&
    \int_{M^{i}_{\rm \HI}}^{M^{i+1}_{\rm \HI}} \phi(M_{\rm \HI}) \, d\log(M_{\rm \HI}) \nonumber \\
    && \times \, \omega D_{H_0} 
    \int_{z_l(M_{\rm \HI})}^{z_u(M_{\rm \HI})} \frac{D^2_C(z)}{E(z)} \, dz 
    \int_{0}^{I_{m}(z)} \sin I \, dI
    \label{eq:count_model}
\end{eqnarray}
where the first integral is the mass function integral, which computes the number density within mass bin $(M^i_{\rm \HI},M^i_{\rm \HI}+dM_{\rm \HI})$. 
The mass bins are uniformly spaced in logarithmic scale over the full mass range 
$(M_{\rm min}, M_{\rm max}) = (10^7, 10^{12})~M_{\odot}$. 
The second integral represents the comoving maximum volume accessible to each source within the detection threshold, bounded by the lower~$z_l$ and upper~$z_u$ redshift limits set by its mass. $D_C(z)$ denotes the comoving distance calculated at the redshift of the source, and $E(z)$ is a function that depends on redshift and the cosmological parameters. Since edge-on \HI sources $(I=90^{\circ})$ are difficult to detect at large distances, the third integral incorporates the impact of inclination on detectability: a source of \HI mass $M_{\rm \HI}$ within the detectable volume can only be observed up to a maximum redshift and maximum inclination angle $I_m$, beyond which it falls below the detection threshold. 
The~\Fref{fig:mz_relation} of~\Aref{ssec:relations}, illustrates how the detection threshold defines the relation between accessible redshifts and inclination. 
With the same approach, we defined the model for the average stacked \HI mass, which is given as,
\begin{eqnarray}
    \langle \boldsymbol{M}\rangle &=&
    \frac{1}{N_s}\int_{M_{min}}^{M_{max}} \, M_{\rm \HI} \,\phi(M_{\rm \HI}) \, d\log(M_{\rm \HI}) \nonumber \\
    && \times \, \omega D_{H_0} 
    \int_{z_l(M_{\rm \HI})}^{z_u(M_{\rm \HI})} \frac{D^2_C(z)}{E(z)} \, dz 
    \int_{0}^{I_{m}(z)} \sin I \, dI 
    \label{eq:stack_model}
\end{eqnarray}
where $N_s$ is the number of sources to be stacked within the full mass range $(M_{\rm min}, M_{\rm max}) = (10^7, 10^{12})~M_{\odot}$ below the detection threshold $(\rm SNR < 5.0)$.
As discussed below, stacking is done with an optical pre-selection, and that has to be folded in the analytical estimation. The pre-selection is incorporated into the analytical estimates using the $\mathrm{M_{\HI}}$--$\rm M_r$ scaling relation and its average scatter, which was derived from the A100-SDSS catalogue.

As the joint optical-\HI survey accounts for the telescope's primary beam sensitivity, our direct detection and average \HI mass models account for this effect by replacing stochastic sampling with an angular space average $ \langle \beta\rangle$, given by
\begin{equation}
    \langle \beta\rangle = \frac{\int_0^{\theta_{max}}\beta(\theta)\sin{\theta}d\theta}{\int_0^{\theta_{max}}\sin{\theta}d\theta}
    \label{eq:expected_beam}
\end{equation}
Here, $\beta(\theta)$ denotes the telescope's primary beam response in a given direction, 
and $\theta_{\rm max}$ is the maximum angle corresponding to the telescope's field of view.

%%%%%%%%%%%%%%%%%%%%%%%% direct detection %%%%%%%%%%%%%%%%%%%%%
\begin{figure*}
    \centering
    \includegraphics[width=0.4\linewidth]{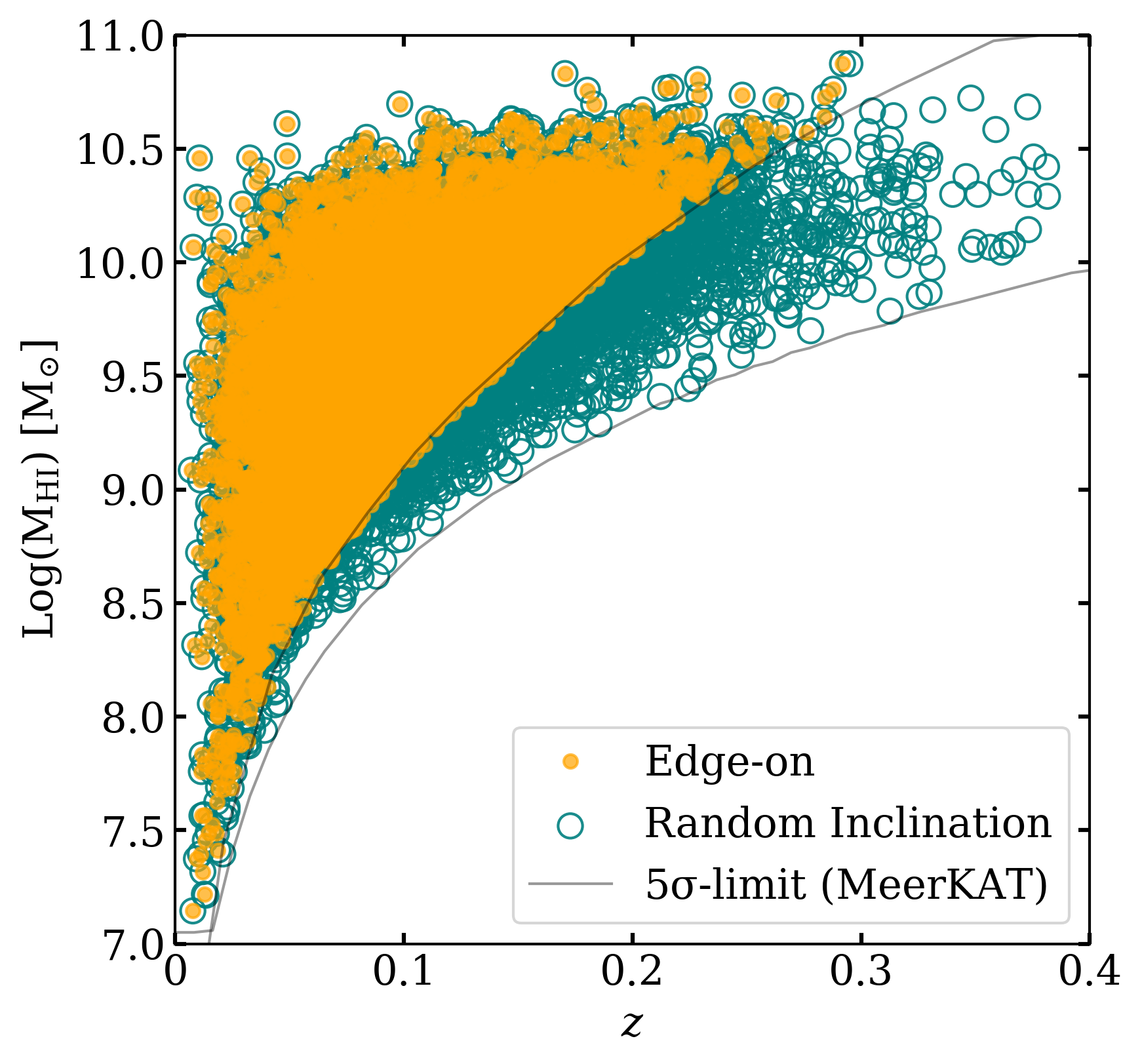}
    \includegraphics[width=0.4\linewidth]{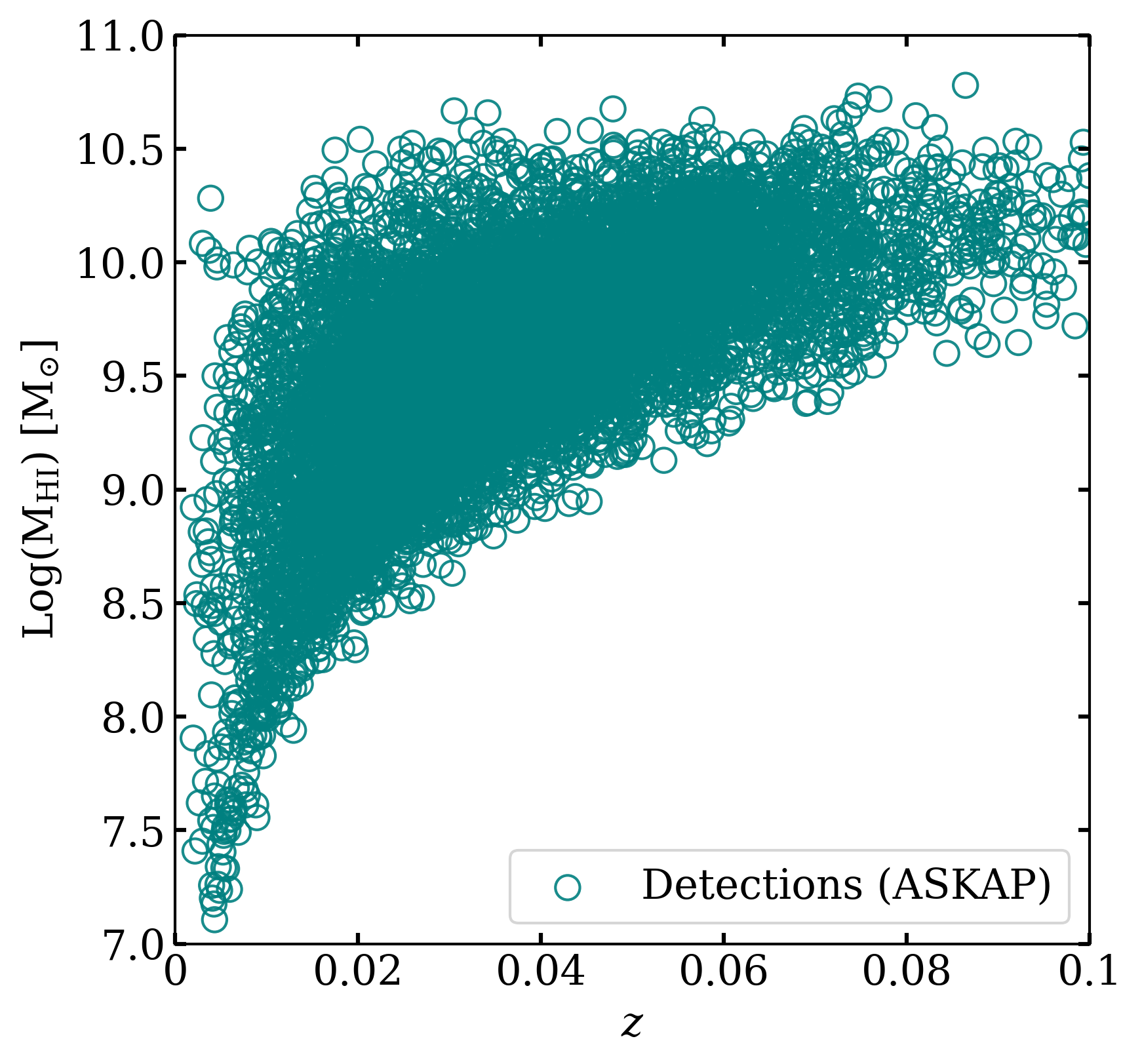}
    \caption{The inclination effect on our model of generating mock catalogue: The scatter points in \emph{left-panel} represent the number of sources detected blindly in the MeerKAT mock survey. The \emph{solid orange circles} indicate the number of galaxies assuming an edge-on orientation, while the \emph{open teal circles} show the detection when random inclinations are incorporated into our simulation. Accounting for random inclinations leads to a significant increase in the expected number of detected sources. Numbers are for the $5\sigma$ flux limit in $20$ deg$^2$ sky area in $25$~hrs of integration per MeerKAT pointing. The solid grey lines mark these flux limits for each case. The \emph{right panel} shows the direct detections from a 6-pointing survey, assuming 8-hrs of ASKAP integration per pointing, with each pointing covering a survey area of $30$ deg$^2$.}
    \label{fig:incl_effect}
\end{figure*}

%%%%%%%%%%%%%%%%%%%%%%%% Counts models for direct detection %%%%%%%%%%%%%%%%%%%%%
\begin{figure*}
    \centering
    \includegraphics[width=0.7\linewidth]{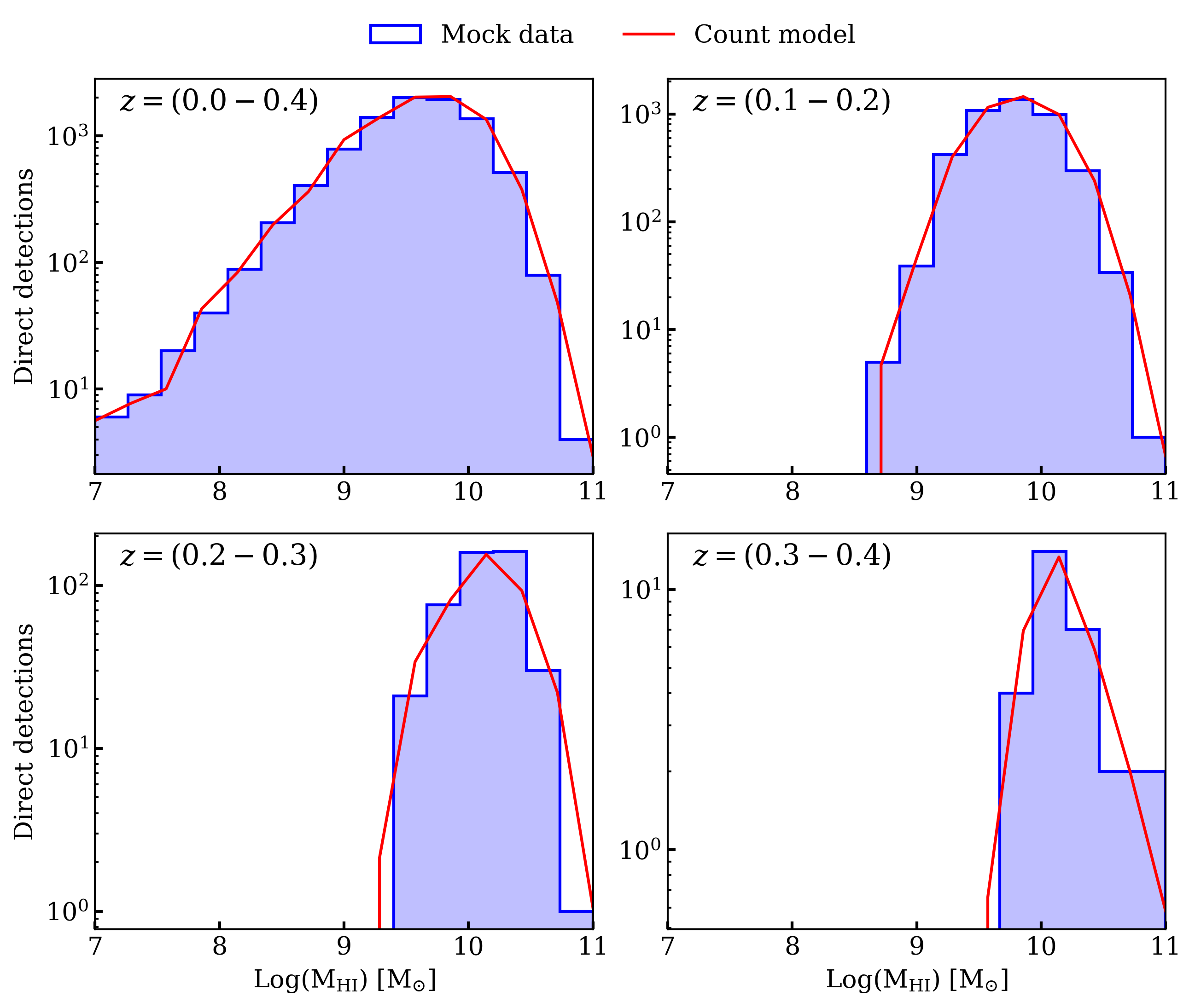}
    \caption{ The distribution of direct detection counts as a function of \HI mass in the joint MeerKAT mock surveys corresponds to an area of $20$ deg$^2$. The histogram of detected sources is shown in \emph{blue}, while the \emph{solid red} line represents the number counts predicted by our analytic model~\Eref{eq:count_model}, corresponding to the number of direct detections. The four panels correspond to different redshift bins, as indicated in the legends, and $0 < z < 0.1$ redshift bin is not shown here.}
    \label{fig:count_model}
\end{figure*}

To determine the posterior probability distribution and fit the \HI mass function for mock observables, we use PyMultinest~\citep[e.g.,][]{Feroz_2009,pymultinest} (Monte-Carlo sampling) to sample the prior parameter space. 
PyMultiNest is a nested sampling algorithm and Bayesian inference tool that computes the evidence and generates posterior samples from the distribution, along with their associated uncertainties. 
In this work, we adopt the median of the posterior samples and their corresponding uncertainties for the three parameters, obtained by maximising the likelihood in fitting the \HIMF. 
In the table \ref{tab:prior}, we list the priors of the parameters used for the parameter sampling and evidence computation.
%%%%%%%%%%%%%%%%%%%%%%%%%%%%%%%%%%%%%%%%%%%%%%%%%%%%%%%%%%%%%%%%%%%%%%%%
\begin{table}
    \centering
    \caption{\justifying The prior range in our model for the mock observable used in the MultiNest runs.}
    \begin{tabular}{rccc}
    \hline \vspace{0.1cm}
    \HIMF Parameters      & Input & Prior Distribution/Range\\
    \hline \vspace{0.1cm}
    $\alpha$ & -1.01   & uniform $\in\,[-5,0]$\\
    $\log_{10}(\rm M^{*}_{\rm HI}/\rm M_{\odot})$&9.80&uniform $\in\,[8,11]$ \\
    $\log_{10}(\phi^{*}/\rm Mpc^{3})$ &  -2.15  & uniform $\in\,[-5,0]$
    \\
    \hline
    \end{tabular}
    \label{tab:prior}    
\end{table}

%%%%%%%%%%%%%%%%%%%%%%%%%%%%%%%%%%%%%%%%%%%%%%%%%%%%%%%%%%%%%%%%%%%%%%%%
\section{Results and Discussion}
\label{sec:result}

We present results for SKA precursor surveys, including MIGHTEE-HI and LADUMA with MeerKAT and WALLABY with ASKAP, using joint optical–\HI mock survey.

\subsection{Direct detections}
\label{ssec:DD}

We simulate joint mock surveys for MeerKAT and ASKAP, covering approximately $1$~deg$^2$ and $30$~deg$^2$ per pointing, respectively. 
We present direct detection results assuming 20 MeerKAT pointings for MIGHTEE-HI ($25$~hrs integration per pointing), covering a survey area of $20$~deg$^2$, a single MeerKAT pointing for LADUMA with a survey area of $1$~deg$^2$ ($1000$~hrs integration per pointing), and 6 ASKAP pointings for WALLABY ($8$~hrs integration per pointing), covering a total survey area of $180$~deg$^2$.  
Radio surveys, like most surveys are flux-limited and detect sources above a certain flux threshold. 
For a fixed total flux, galaxies with narrow linewidth (face-on systems) having higher peak flux densities are more likely to be detected than those with broad linewidth (edge-on systems). 
We also present estimates for a model that neglects inclination effects by assuming all sources are edge-on, to examine how the detection threshold varies in the mass-redshift plane. 
The impact of inclination is illustrated in \emph{left panel} of~\Fref{fig:incl_effect}, for the MeerKAT mock survey. 
The \emph{orange solid circles} are direct detections $(\rm SNR>5)$ in the mass-redshift space corresponding to a model where all sources are edge-on. 
In contrast, the \emph{teal-blue open circles} show the sources corresponding to the model where we randomly assign the inclination $I$ such that the~$\cos{(I)}$ is distributed within $[0,1]$. 
Taking into account random inclinations in the model significantly increases the number of detected sources. 
The grey lines indicate the $5\sigma$ detection limits for the MeerKAT sensitivity for both models. 
Therefore, accounting for inclination effects is essential for unbiased radio survey predictions and for the accurate interpretation of observations. The right panel displays the directly detected sources in the mass-redshift plane for the ASKAP mock surveys, assuming a model that accounts for random inclinations.

Here we also present an analytic model of the source counts \Eref{eq:count_model} for direct detections in the MeerKAT mock survey. In ~\Fref{fig:count_model}, different panels show the distribution of the number of direct detections as a function of \HI mass at different redshifts. 
The \emph{top-left} panel shows the distribution of the sources directly detected in the joint mock surveys (in \emph{blue}) for the broad redshift bin $0 < z < 0.4$ (global), while the \emph{red curve} shows the number counts (direct detections) predicted by the analytic model~\Eref{eq:count_model}. 
The other panels show the same distribution for different redshift bins; $0 < z < 0.1$ redshift bins is not shown here. 
The number of direct detections decreases with increasing redshift, with only the most massive sources being detectable at the farthest edge of the survey. 
Therefore, measuring the \HI mass function through direct detections at the farthest edge of the survey (below $\sim10^{9.8}~M_{\odot}$ in this case) is challenging. 

%%%%%%%%%%%%%%%%%%%%%%%%%%%%%%%%%%%%%%%%%%%%%%%%%%%%%%%%%%%%%%
\begin{figure*}
    \centering
    \includegraphics[width=0.45\linewidth]{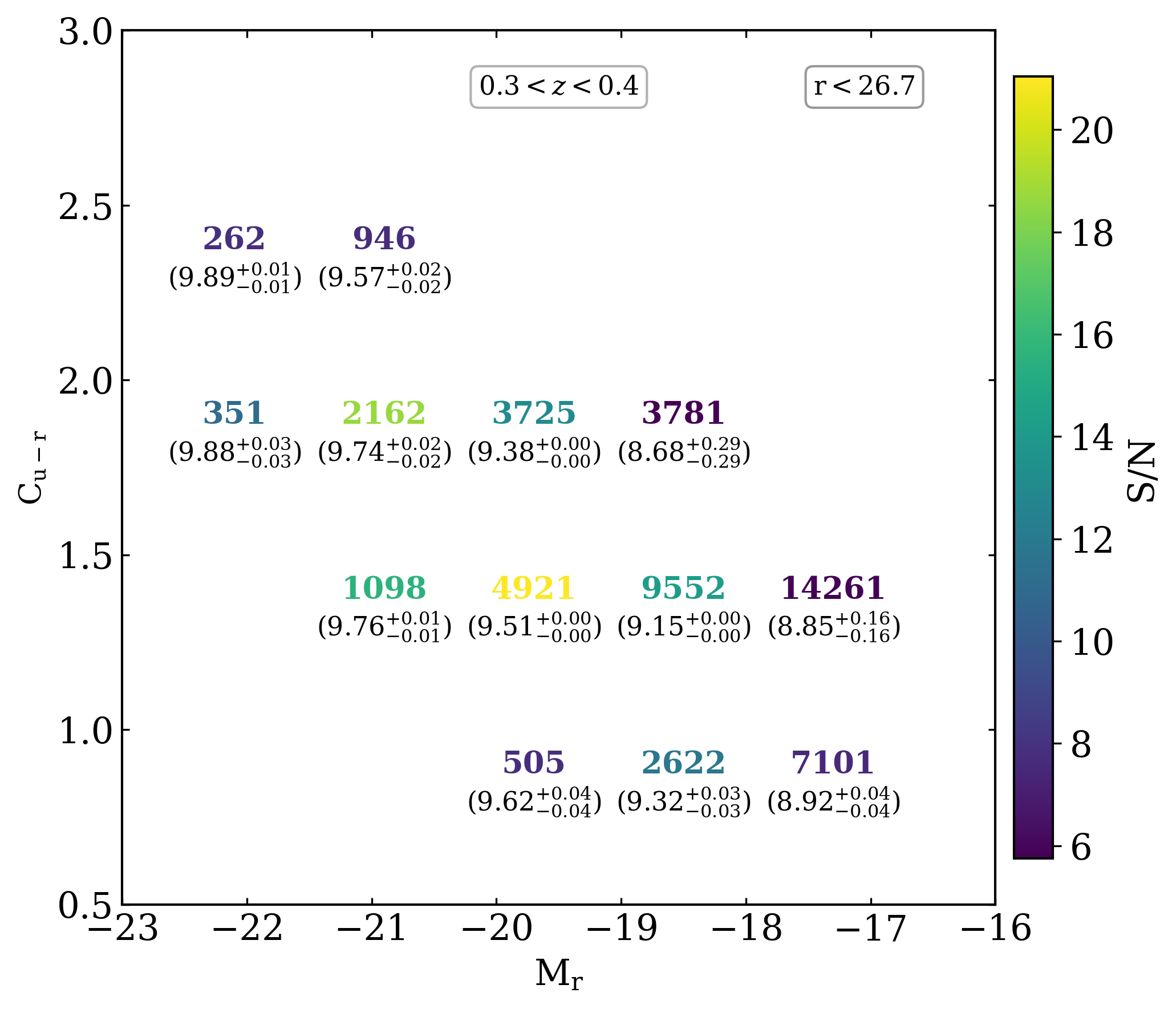}
    \includegraphics[width=0.45\linewidth]{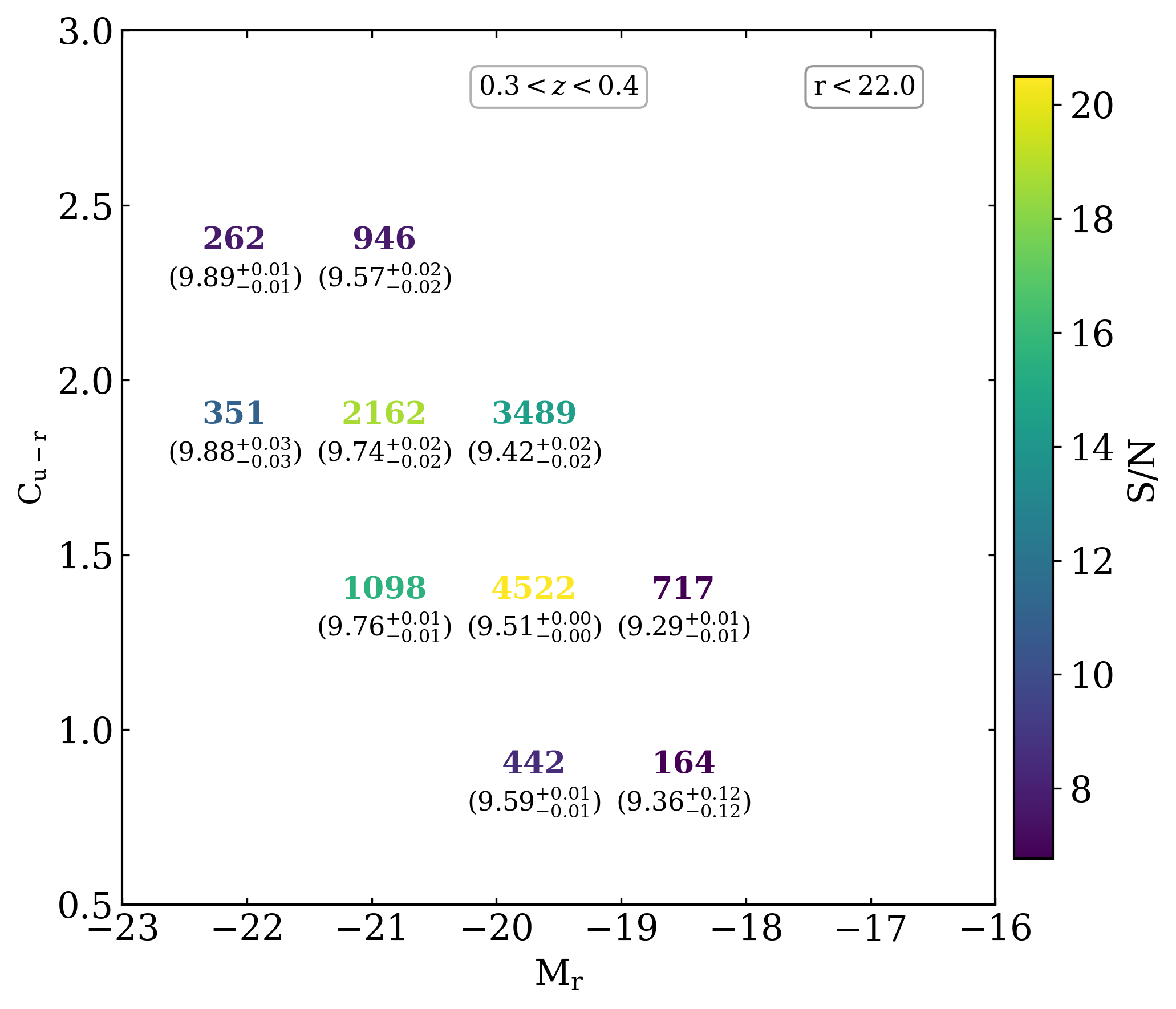}
    \caption{Stacking significance detection: This figure shows the stacking significance of sub-threshold sources in the redshift range $0.3 < z < 0.4$ for a MeerKAT mock survey covering $4~\mathrm{deg}^2$, evaluated across the various colour-magnitude pixels. Only pixels with stacking detections above $5\sigma$ are shown. The \emph{left panel} presents the number of stacked \HI (COSMOS cut $r<26.7$) spectra within pixels of width $\Delta C_{ur} = 0.5$ in colour and $\Delta M_r = 1.2$ in magnitude. The number of spectra stacked in each pixel is colour-coded according to the stacking significance (SNR). The mean \HI mass along with its $1\sigma$ uncertainties is also displayed in the corresponding pixel. The uncertainties are calculated over $50$ realisations. The right panel shows the corresponding result for an optical r-band cut $r<22$.}
    \label{fig:stack_pix}
\end{figure*}

%%%%%%%%%%%%%%%%%%%%%%%%%%%%%%%%%%%%%%%%%%%%%%%%%%%%%%%%%%%%%%%%%%%%%%%%%%
\begin{table*}
    \centering
    \caption{\justifying Number of stacked spectra $N_s$, their average \HI mass $\langle M_{\rm \HI}\rangle$, and the stacking detection significance (SNR): Columns (2)--(5) correspond to the four redshift bins of the MIGHTEE-\HI\ survey covering $4~\mathrm{deg}^2$. The uncertainties $\sigma_{\langle M_{\rm \HI}\rangle}$ in the average \HI mass are derived from 50 realisations. Columns (6) and (7) show these observables for the two redshift bins of the LADUMA survey covering $1~\mathrm{deg}^2$. Column (8) presents the corresponding  results for a single ASKAP pointing covering~$30~\mathrm{deg}^2$.}
    \begin{tabular}{rcccc|cc|c}
    \hline
          &MIGHTEE-\HI ($4$ deg$^2$)       &       &    &     & LADUMA ($1$ deg$^2$) & &WALLABY ($30$ deg$^2$)  \\
    \hline
    Observables  &0 < $z$ < 0.1  &0.1 < $z$ < 0.2    & 0.2 < $z$ < 0.3                & 0.3 < $z$ < 0.4  & 0.4 < $z$ < 0.5 & 0.5 < $z$ < 0.6 & 0 < $z$ < 0.1\\
    \hline

    $N_{s}$       &  $2010$ & $17387$   & $45023$  &  $79057$  & $28836$  & $39100$ & $24454$   \\

    $\log_{10}\big(\frac{\langle M_{\rm \HI}\rangle}{M_{\odot}} \big)$       &  $8.27\pm0.01 $ &  $8.85\pm 0.02$  & $8.99\pm 0.015$  &  $9.01\pm0.01$  & $8.94\pm 0.001$  & $8.98\pm 0.01$ & $8.95\pm 0.07$  \\

    SNR(Stacking)    & $30.9$  &   $45.68$ & $32.45$  & $22.08$   & $43.95$  & $32.22$ & $39.9$ \\
    
    \hline
    \end{tabular}
    \label{tab:observables}    
\end{table*}  
%%%%%%%%%%%%%%%%%%%%%%%%%%%%%%%%%%%%%%%%%%%%%%%%%%%%%%%%%%%%%%%%%%%%%

\subsection{Indirect detections: \textsc{\HI} stacking}
\label{ssec:indirect_det}

For \HI sources which are below the detection level, we estimate the significance of detection via stacking. 
Following the stacking technique discussed in the ~\Sref{ssec:ska_prec}, we present the stacking significance results in~\Tref{tab:observables}. 
Spectroscopic redshifts for sources are required for stacking, and spectroscopic surveys typically have a flux limit in the optical or near IR.  
To emulate this in our mock catalogues, we use an optical selection for stacking.  
Using the COSMOS apparent magnitude cut in the i-band ($i < 26.7$), we select the galaxy population within a given redshift bin and perform \HI stacking, with jitter added to emulate observational uncertainties. 
Note that for the stacking, we use the survey areas of $4~\mathrm{deg^2}$, $1~\mathrm{deg^2}$, and $30~\mathrm{deg^2}$ for MIGHTEE-\HI, LADUMA, and WALLABY, respectively, to present the statistical significance. 
Across four redshift bins (MIGHTEE-\HI), the statistical significance of stacking exceeds $\gtrsim 20\sigma$, indicating that the average \HI mass can be measured with high significance. 
A similar level of stacking significance can be achieved in the LADUMA and WALLABY survey areas. 
With $\sim10^4$ galaxies in the $(0.3<z<0.4)$ bin, one can divide the sample across colour–magnitude cells and determine the average \HI mass in each cell separately, enabling the measurements of average properties of galaxies. 
Such a determination allows us to compare the properties of galaxies across redshifts and carry out a direct comparison with the low redshift sample. 
In \emph{left-panel} of~\Fref{fig:stack_pix}, we show subsets of these sources and their stacking ($r<26.7$) significance across the color–magnitude plane. In this plot, we examine how accurately the average \HI\ mass can be recovered, given the survey sensitivity and a uniform spatial distribution of sources across the field of view. The number of spectra stacked in each pixel is colour-coded according to the stacking significance (SNR). The average \HI mass recovered via stacking is also displayed in the corresponding pixel. Note that the $1\sigma$ uncertainty in the stacked \HI mass includes: (i) bias in recovering the true input mean mass through stacking, (ii) statistical error estimated from 50 realisations. The total quoted error (bias + statistical) increases for faint sources (with $M_r > -18$), even when the number of such stacked spectra is very large, because their spectra are much noisier. With the stacking of such faint sources, the improvement in the co-added signal is not significant. This missing flux propagates through the stacking, limiting its ability to accurately recover the average \HI mass with high significance. In the \emph{right panel}, we also present the stacking significance (stacking SNR) obtained using a shallower apparent magnitude cut of $r <22 $.

The average \HI properties of galaxies can be connected to their optical properties in regions of the color–magnitude plane where the stacking SNR is $\gtrsim 8$, providing a framework to investigate the evolution of \HI gas (e.g., \HIMF, optical to \HI scaling relations) across redshift and within the colour–magnitude diagram. For instance, we demonstrate the strength of employing \HI stacking to estimate the \HI mass function in the next section.

%%%

%%%%%%%%%%%%%%%%%%%%%%%%%%%%%%%%% Constraints %%%%%%%%%%%%%%%%%
\begin{figure*}
    \centering
    \includegraphics[width=0.49\linewidth]{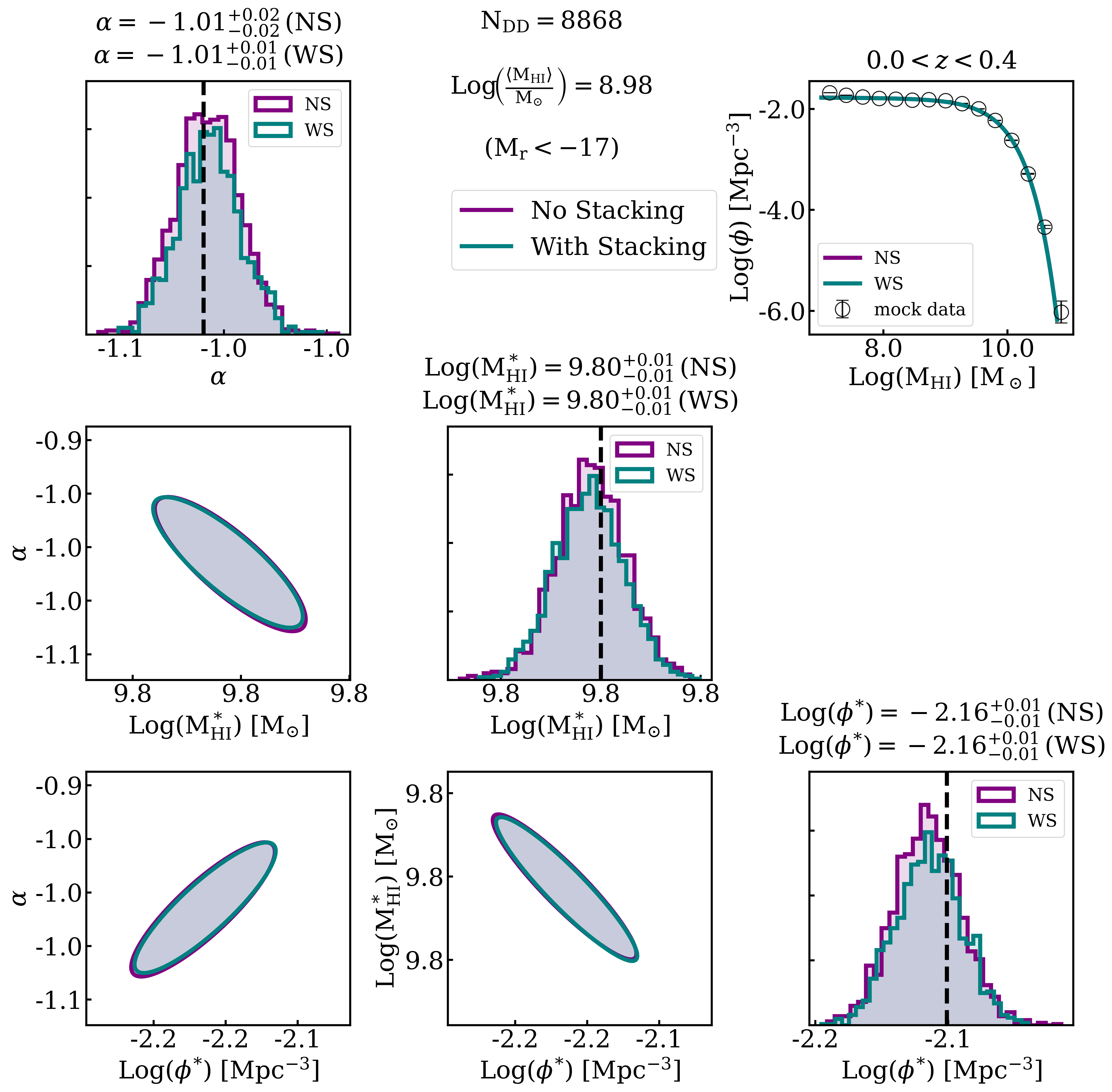}
    \includegraphics[width=0.49\linewidth]{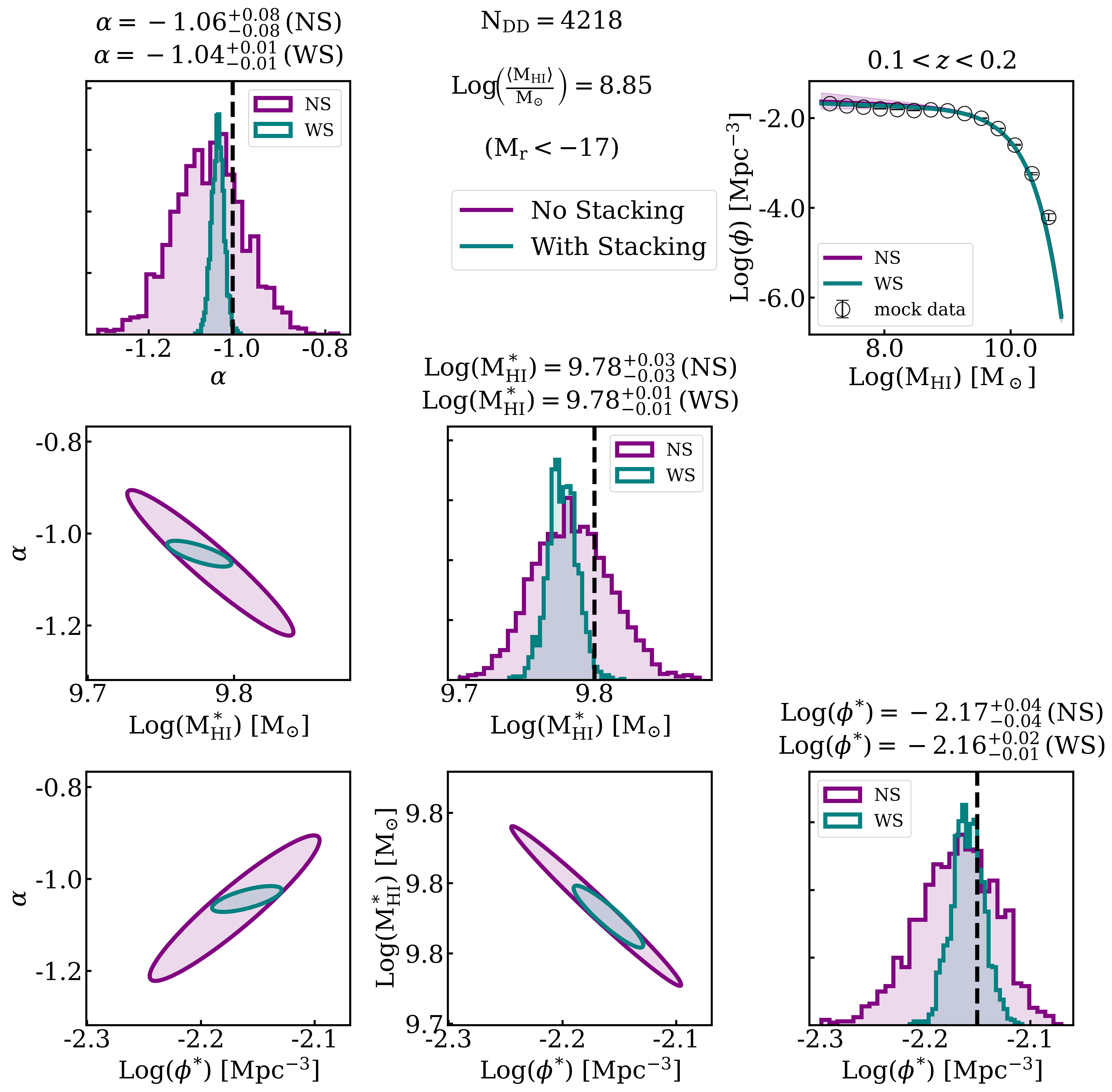}
    \caption{ Posterior probability distribution for the three parameters of \HI mass function: \emph{Left panel} shows the corner plots for the three parameters derived using the Bayesian method for redshift $0 < z < 0.4$. Two cases are considered: (i) using only direct detections, no stacking and (ii) including stacking measurements in the likelihood along with direct detections. The corresponding posterior distributions are shown in \emph{magenta} and \emph{teal} color, respectively. Instead of the actual distribution of points, we show $2\sigma$ error ellipses for each case. The median values of the posterior corresponding to the \HIMF parameters are shown on top of the distribution with: No stacking~(NS) and with stacking~(WS). The top-middle lists the number of direct detections ($N_{\mathrm{DD}}$) and the average stacked \HI mass 
    $\langle M_{\rm \HI}\rangle$ for sources brighter than $M_r < -17$. The top-right small panel shows the \HIMF plots as a function of \HI mass for both cases, along with estimates from mock data points (black open circles). The small panel also shows the \HI mass function with uncertainties corresponding to the $68\% $ posterior range, though the shaded regions are barely visible due to the very small uncertainties in the parameters. \emph{Right panel} shows the posterior distributions, corner plots and \HIMF plots for the redshift range $0 < z < 0.1$.}
    \label{fig:mkat_himf1}
\end{figure*}
%%%%%%%%%%%%%%%%%%%%%%%%%%%%%%%%%%%%

%%%%%%%%%%%%%%%%%%%%%%%%%%%%%%%%%%%%
\begin{figure*}
    \centering
    \includegraphics[width=0.49\linewidth]{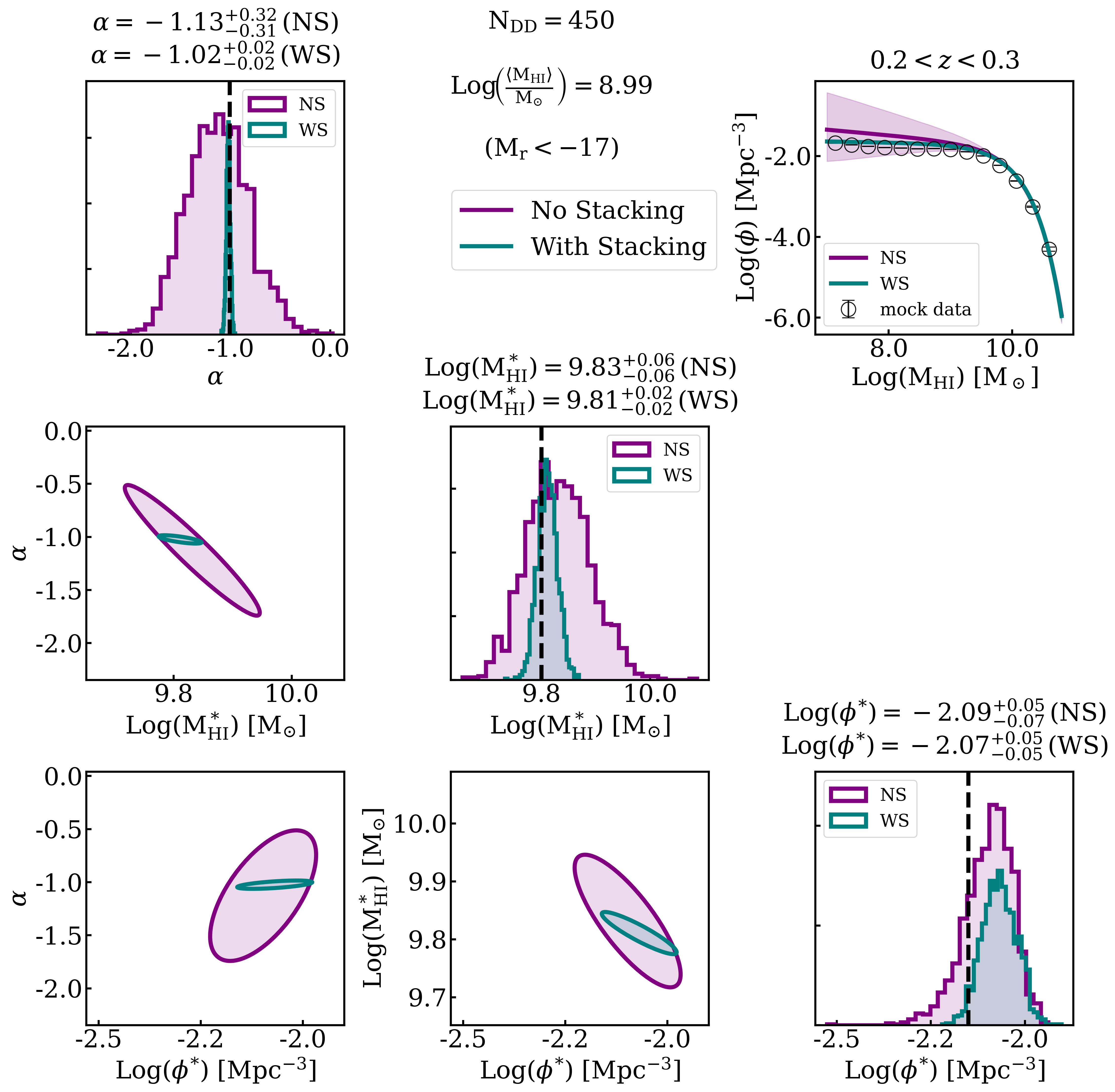}
    \includegraphics[width=0.49\linewidth]{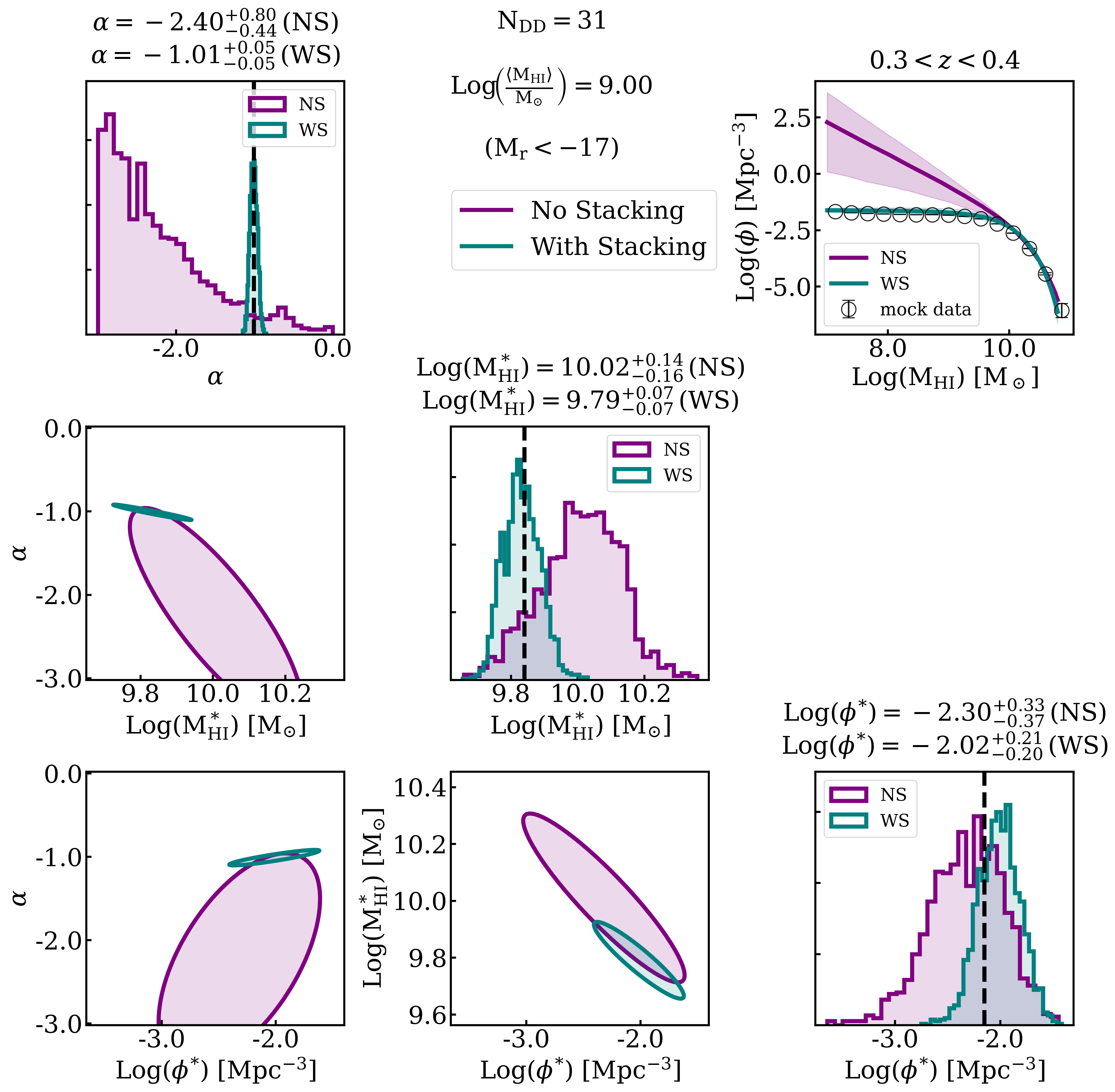}
    \caption{ Posterior probability distributions for the three parameters of the \HI mass function in the redshift ranges $0.2 < z < 0.3$ (\emph{left set of 7 sub-panels}) and $0.3 < z < 0.4$ (\emph{right set of 7 sub-panels}). At these redshift bins, $68\%$ \emph{magenta shaded region} is visible in the \emph{top-right small panel} for the direct detection (no stacking; NS) case, as the uncertainty is larger. Reader is referred to Fig.~\ref{fig:mkat_himf1} for a detailed explanation of the various panels and sub-panels.}
    \label{fig:mkat_himf2}
\end{figure*}

\subsection{Constraints on \textsc{\HIMF} parameters}
\label{ssec:himf_constrain}

 In this section, we focus on reconstructing \HIMFs from joint mock observations. 
 We present the \HIMF estimation results derived from the MIGHTEE-\HI surveys using Bayesian analysis. 
 We divide the mock data into four redshift bins to explore how \HI mass function constraints vary with redshift. 
 Using the assumed model of the \HI\ mass function (Eq.~\ref{eq:himf}) and the number counts in mass bins (Eq.~\ref{fig:count_model}), we constrain the three parameters: $\alpha$, $\log_{10}(\rm M_{\rm HI}/M_{\odot})$, and $\log_{10}(\phi^{*})$. 
 The prior range corresponding to three parameters is given in the table~\ref{tab:prior}. 
 The bin-wise number counts break the degeneracy among these parameters. 
 The posterior probability distribution of the parameters is shown in~\Fref{fig:mkat_himf1}. 
 The \emph{left} set of seven \emph{sub-panels} (showing corner plots and fitted \HIMFs) is for the redshift range $0<z<0.4$. 
 In the sub-panels, the posteriors obtained using direct detections only (no stacking; NS) are shown in \emph{magenta} color, whereas the \emph{teal} color represent the posteriors obtained from the combined likelihood (with stacking; WS) that includes both direct detections and \HI stacking. 
 \HI stacking is performed for all sources (brighter than $M_r = -17~\mathrm{ and}~r<26.7 )$ within $4~\mathrm{deg^2}$ sky area. 
 The direct detections are considered across all pointings assumed for a given survey. 
 The ellipses in the corner plots denote the $2\sigma$ uncertainties around the best-fit values (median of the posteriors). 
 The \emph{dashed black} vertical lines indicate the corresponding input parameter value used in the joint mock survey. 
 The best-fit parameters of the reconstructed \HIMF are consistent with the input values within the uncertainties. 
 The best-fit values for both cases, with and without stacking, are displayed at the top of each sub-panel. 
 At the \emph{top middle}, the number of direct detections $N_{DD}$ and the average stacked \HI mass $\langle M_{\rm \HI}\rangle$ for sources with $M_r < -17$ are shown. 
 It can be seen from the corner plots and \HI mass function (\emph{top-right sub-panel}) plots that there is no significant difference between the \HIMFs estimates obtained with (WS) and without stacking (NS) over the global(broad) redshift range, as enough galaxies are directly detected across the entire mass range (see \emph{top-left} panel of~\Fref{fig:count_model}).

 On the other hand, in the \emph{right} set of seven \emph{sub-panels} of~\Fref{fig:mkat_himf1}, corresponding to the redshift range $0.1 < z < 0.2$, a significant difference can be seen between the two estimates: With and without stacking. 
 Note that combining the stacking measurements with direct detections significantly reduces the parameter uncertainties, which would not be achievable if only direct detection were included in this redshift range. 
 The estimated parameters are consistent with the input values within the $3\sigma$ confidence level. 
 For the redshift range $0.2 < z < 0.3$, as seen in the \emph{bottom-left} panel of~\Fref{fig:count_model}, the number of direct detections is smaller and limited to higher masses. 
 %%%%%%%%%%%%%%%%%%%%%%%%%%%%%%%%%%%%%%%%%%%%%%%%%%%%%%%%%%%%%%%%%%%%%%%%
\begin{table*}
    \centering
    \caption{\justifying Parameters of the reconstructed \HI mass function from MeerKAT's mock survey with the area of~$20$~deg$^2$. These parameters are estimated using mock observables: Number of direct detection and stacked average \HI mass of galaxies ($\rm M_{r} < -17.0$) in FoV.}
    \begin{tabular}{rccccc|ccc}
    \hline
          &       &       &  MIGHTEE-\HI ($20$ deg$^2$)  &    & & LADUMA ($1$ deg$^2$) &  \\
    \hline
          Parameter  & 0 < $z$ < 0.4  &0 < $z$ < 0.1         &0.1 < $z$ < 0.2 & 0.2 < $z$ < 0.3  & 0.3 < $z$ < 0.4  & 0.4 < $z$ < 0.5 & 0.5 < $z$ < 0.6\\
    \hline   
    $\alpha$  & $-1.01^{+0.01}_{-0.01}$ &  $-1.01 ^{+0.01}_{-0.01}$ & $-1.04^{+0.01}_{-0.01}$    &  $-1.02^{+0.02}_{-0.02}$       &  $-1.01^{+0.05}_{-0.05}$  & $-1.0^{+0.01}_{-0.02}$ & $-1.02^{+0.02}_{-0.02}$\\

    $\log_{10}(\rm M^{*}_{\rm HI}/\rm M_{\odot})$           & $9.80^{+0.01}_{-0.01}$ &  $9.80^{+0.01}_{-0.01}$ & $9.79^{+0.01}_{-0.01}$    &  $9.81^{+0.02}_{-0.02}$ &  $9.79^{+0.07}_{-0.07}$  & $9.80^{+0.02}_{-0.02}$ & $9.82^{+0.03}_{-0.02}$\\

    $\log_{10}(\phi^{*})$          & $-2.16^{+0.01}_{-0.01}$ & $-2.17^{+0.01}_{-0.01}$ & $-2.16^{+0.02}_{-0.01}$   &  $-2.07^{+0.05}_{-0.05}$     &  $-2.02^{+0.21}_{-0.20}$  & $-2.10^{+0.04}_{-0.04}$ & $-2.04^{+0.08}_{-0.09}$\\
    
    \hline
    \end{tabular}
    \label{tab:himf_values_mkat}    
\end{table*}  
%%%%%%%%%%%%%%%%%%%%%%%%%%%%%%%%%%%%%%%%%%%%%%%%%%%%%%%%%%%%%%%%%%%%%%%%
%%%%%%%%%%%%%%%%%%%%%%%%%%%%%%%%%%%%%%%%%%%%%%%%%%%%%%%%%%%%%%%%%%%%%%%%
\begin{table}
    \centering
    \caption{\justifying Parameters of the reconstructed \HI mass function from ASKAP's mock survey with the area of~$30$~deg$^2$. These parameters are estimated using mock observables: Number of direct detection and stacked average \HI mass of galaxies ($\rm M_{r} < -17.0$) in FoV.}
    \begin{tabular}{rcc}
    \hline
    Surveys      &WALLABY-\HI ($30$ deg$^2$)       &\\
    \hline
     HIMF Parameter      & 0 < $z$ < 0.1 &   \\ 

     $\alpha$   & $-0.99^{+0.05}_{-0.04}$ \\

     $\log_{10}(\rm M^{*}_{\rm HI}/\rm M_{\odot})$ & $9.81^{+0.02}_{-0.02}$ \\

     $\log_{10}(\phi^{*})$  & $-2.13^{+0.03}_{-0.02}$ \\
    \hline
    \end{tabular}
    \label{tab:himf_values_akap}    
\end{table}  

%%%%%%%%%%%%%%%%%%%%%%%%%%%%%%%%%%%%%%%%%%%%%%%%%%%%%%%%%%%%%%%%%%%%%%%%

 Therefore, in the \emph{left} panel of ~\Fref{fig:mkat_himf2}, the \HI mass function parameters are not well constrained and show large uncertainties when considering direct detections only (shown in \emph{magenta} color posteriors) and $68\%$ shaded region is clearly visible in the small \HIMF panel. Accounting for stacking not only recovers the input parameters more accurately but also significantly reduces the uncertainties, as evident from the \emph{teal} color posterior distributions. 
 The right set of seven \emph{sub-panels} in~\Fref{fig:mkat_himf2} shows the posterior distribution of the parameter for last redshift bin ($0.3 < z < 0.4$) of the MeerKAT surveys. 
 In this case, the number of direct detections is limited to only a few bins at the higher-mass end (see \emph{bottom-left} panel of ~\Fref{fig:count_model}), resulting in large uncertainties in parameters of reconstructed \HIMF. 
 When \HI stacking is included, the posterior distributions become narrower, and the corresponding parameter estimates are consistent with the input values adopted in our joint mock survey. 

 We present the parameters of reconstructed \HIMFs from combined stacking with direct detections in Table~\ref{tab:himf_values_mkat} and Table~\ref{tab:himf_values_akap} for the MeerKAT and ASKAP surveys, respectively. 
 From these estimates, our simulations show that stacking measurements when combined with direct detection are capable of constraining the \HIMF with high significance, even at the boundary of \HI surveys.

\subsection{Impact of clustering and beam confusion}
\label{ssec:clustering_impact}
In the absence of spatial correlation (galaxy clustering), our mock catalogues do not capture field-to-field fluctuations(cosmic variance) or confusion caused by multiple galaxies within the same resolution element and channels caused by small scale clustering. Here, we discuss the potential impact of its absence, along with the effects of beam confusion.  Although, we do not expect our mock catalogue or the derived \HIMF results to be significantly affected for large volume surveys. However, in real pencil beam surveys (LADUMA-like), these fluctuations can increase the variance in recovered quantities. In order to estimate the expected cosmic variance, we follow the formalism discussed in~\citet{Driver_2010}. For survey area and redshift range considered here, we expect the cosmic variance to be $\sim 5\text{--}6\%$ for MIGHTEE-HI in each redshift bin ($\Delta z = 0.1$), $\sim 15\%$ for WALLABY, and $\sim 25\text{--}26\%$ for LADUMA in each redshift bin. Clustering may lead to source blending due to correlated companions falling within the synthesized beam and velocity window. Following~\citet{Duffy2012}, we perform a simple confusion estimate by assuming a correlation function. For WALLABY, we find a confusion rate of $\sim 4\%$. In contrast, the relatively high angular and spectral resolution of MeerKAT significantly reduces confusion ($<1\%$) for MIGHTEE-HI and LADUMA-like surveys. This implies that confusion-induced bias is likely subdominant compared to cosmic variance. The uncertainties quoted in \HIMF parameters represent lower limits. As discussed earlier, combining stacking significantly reduces the uncertainty in the \HIMF measurement. For the MIGHTEE-HI survey, the \HIMF can therefore be constrained with high statistical significance, despite the effects of cosmic variance and source confusion.

\section{Conclusions}
\label{sec:conclusions}

The neutral hydrogen content of galaxies, which serves as the primary fuel for future star formation, plays a key role in governing their evolutionary pathways. 
Our understanding of galaxy evolution is limited by the lack of direct \HI detections of galaxies at intermediate to high redshifts ($z \gtrsim 0.4$). 
Using direct detections in combination with stacking allow us to constrain the \HI mass function using very few observables out to these redshifts, therefore facilitating the investigation of galaxy evolution at high redshifts. 
In the current work, we create joint optical-\HI catalogues using a novel method and conduct a joint mock survey for the upcoming/ongoing Square Kilometre Array precursors to investigate the expected studies on galaxy evolution at radio and optical wavelengths. 
Using the optical attributes of our mocks, we perform \HI~stacking to estimate the significance of the detection and the average \HI mass of the sources and combine these with direct detections to constrain the mass function. 
Our mock catalogues are simple and unique in their ability to realistically reproduce the properties of galaxies observed in actual surveys (large surveys) at optical and radio wavelengths. 
Our key findings are as follows:

\begin{itemize}

    \item We have developed a novel method for creating the joint optical-\HI catalogue based on the local observations. 
    To validate our technique, we created a joint catalogue with the sensitivity of the ALFALFA survey and compared its global \HI mass function and the distribution of \HI mass in the color magnitude plane with the A100-SDSS observational data. 
    \HI mass distribution in the color magnitude plane and the shape of the \HI mass function in our mock were found to be consistent with that of the A100-SDSS data; this indicates that our method effectively captures the underlying correlation between \HI and optical properties.

    \item We perform a joint mock survey using the estimated correlations between optical and \HI properties of galaxies, taking into account the sensitivity limits of the SKA precursor instruments. 
    We present predictions for the number counts of direct \HI detections, showing how the orientations of galaxies have a significant effect, in contrast to a model that assumes all sources as edge-on. 
    Accounting for inclination increases the predicted number counts for direct detections, as face-on sources (with their sharper peak flux densities) are more likely to be detected than edge-on sources at a given instrumental sensitivity. 
    Accounting for inclination effects is crucial to ensure unbiased predictions from radio surveys and enable an accurate interpretation of observational data.

    \item Optical properties (rest frame U and R band magnitudes) of sources in our joint mock catalogue enable the optical selection and \HI stacking. 
    We simulate the \HI spectral lines, including realistic survey noise for SKA precursor instruments, and perform \HI stacking to estimate the detection significance and the average stacked \HI mass of the selected galaxies. 
    Such estimates cannot be obtained from an \HI-only catalogue. 

    \item For the observables in our joint mock surveys, the number of direct detections and the average stacked \HI mass, we define analytic models: (1) Source counts model for direct detections and (2) Model for the average \HI mass. 
    These models account for all factors used in the mock surveys, including telescope sensitivity, inclination effects, and the primary beam response. 
    The analytic model predicts the expected number of direct detections and mean \HI mass, providing a deterministic reference for comparison with the mock results. 
    With the source-count model for direct detections, we reconstruct the \HI mass function in both a single broad (global) redshift bin and in individual redshift bins using the PyMultinest Monte Carlo sampling. 
    In our mock surveys, we find that the \HIMF can be tightly constrained through the number of direct detections in the global (broad) redshift bin and in the low-redshift bins for both the MIGHTEE-\HI and WALLABY surveys.
    
    \item At higher redshift bins ($z \gtrsim 0.1$), as in the case of MIGHTEE-\HI and LADUMA, the number of direct detections becomes limited and biased toward the high-mass end. We demonstrate that at higher redshift where direct detections are limited and biased, \HIMF can be reconstructed very well by combining \HI stacking measurements with direct detections. Combining stacking with direct detections not only recovers the input \HIMF accurately but also significantly reduces the uncertainties in the inferred parameters, enabling a robust measurement of the expected evolution (if any) of the \HIMF across the redshift range.

    \item Given the tight constraints on the \HIMF using just two inputs, we propose that this approach can be used in future surveys to set priors for the full determination of the \HIMF.
    
\end{itemize}

Our joint mock catalogues can be further improved by accounting for clustering, making them more suitable for predicting outcomes of small-volume (or subsets of large surveys) \HI surveys. 
In our current work, we use the limited sample (A100–SDSS) to establish the correlation between \HI and optical properties of galaxies. 
As a result, the inferred correlations carry uncertainties in a few sparsely populated pixels in the colour–magnitude plane. 
Future radio surveys \citep[SKA;][]{Blyth_2015} and optical surveys \citep[COSMOS-Web, DESI;][]{Casey_2023, Hahn_2023} will measure the \HI\ and optical properties of galaxies with far greater completeness, providing significantly larger samples in both source numbers and redshift coverage. These datasets will reduce the uncertainties and produce more robust joint mock catalogues.

%%%%%%%%%%%%%%%%%%%%%%%%%%%%%%%%%%%%%%%%%%%%%%%%%%%%%%%%%%%%%%%%%%%%%%%%
\section*{Acknowledgements}

The authors acknowledge the use of the PARAM Smriti at NABI Mohali and the HPC facility at IISER Mohali for providing computational resources. This research has made use of NASA's Astrophysics Data System Bibliographic Services. SB thanks the Department of Science and Technology (DST), Government of India, for financial support through the Council of Scientific and Industrial Research-UGC research fellowship. The authors also thank Nishikanta Khandai, Saili Dutta and Tanya Tripty for the useful discussion. The authors also acknowledge Tanya Tripty for deriving photometric properties of ALFALFA sources in SDSS optical bands. This work utilizes ALFALFA Extragalactic Source catalogue with Sloan Digital Sky Survey (SDSS). We thank all members of ALFALFA and SDSS team for their contribution.

%%%%%%%%%%%%%%%%%%%%%%%%%%%%%%%%%%%%%%%%%%%%%%%%%%%%%%%%%%%%%%%%%%%%%%%%
\section*{Data Availability}

All the data products used in this article are publicly available. Additional data products created in the current work can be easily generated using the methods discussed in the text.

%%%%%%%%%%%%%%%%%%%%%%%%%%%%%%%%%%%%%%%%%%%%%%%%%%%%%%%%%%%%%%%%%%%%%%%%
\bibliographystyle{aasjournal}
\bibliography{reference}

%%%%%%%%%%%%%%%%%%%%%%%%%%%%%%%%%%%%%%%%%%%%%%%%%%%%%
\appendix
\section{}
\subsection{Fitting luminosity function and colour}
\label{ssec:fit_lum_clr}

\begin{figure*}
\centering
\includegraphics[width=.32\textheight]{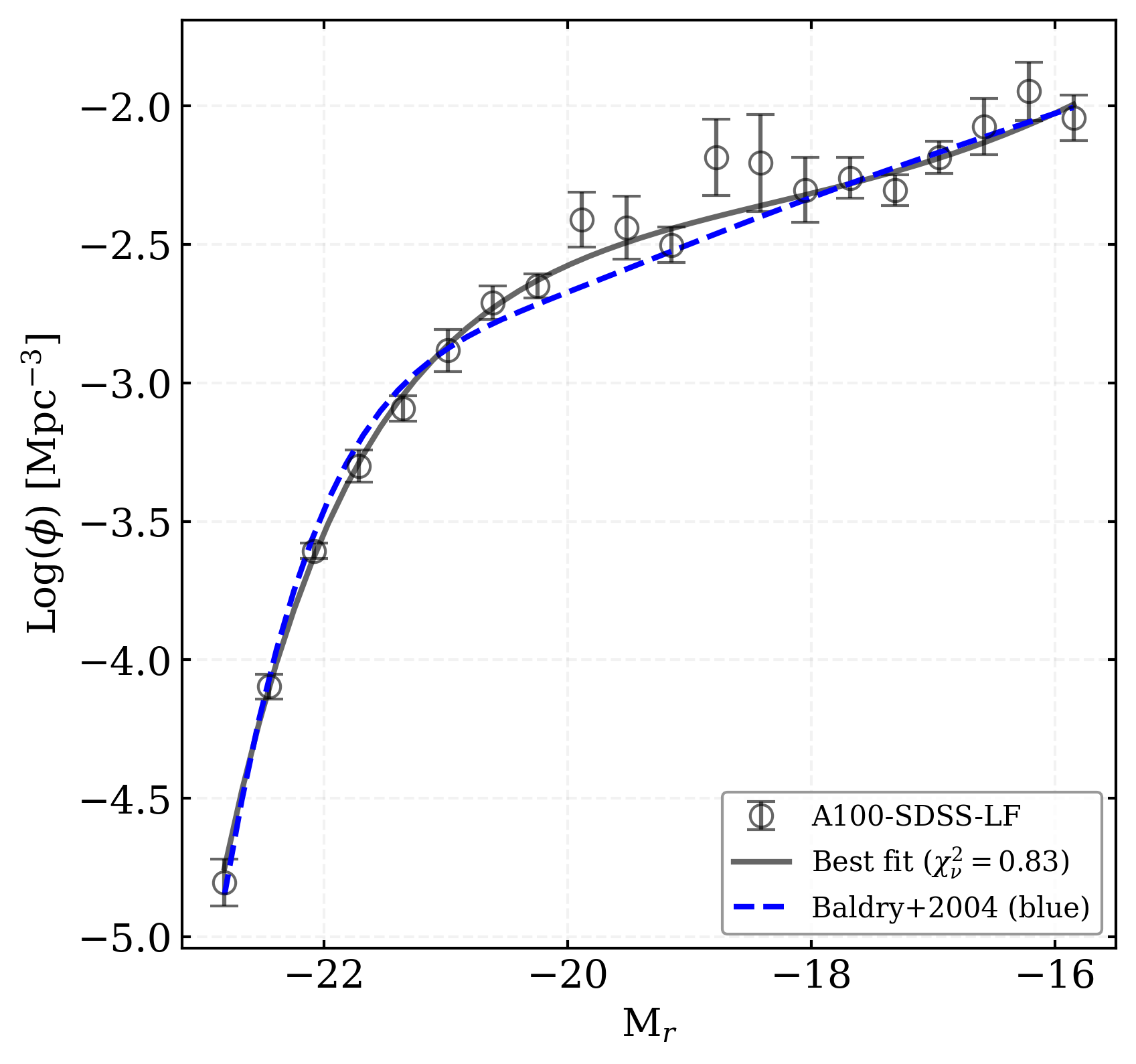}
\includegraphics[width=.33\textheight]{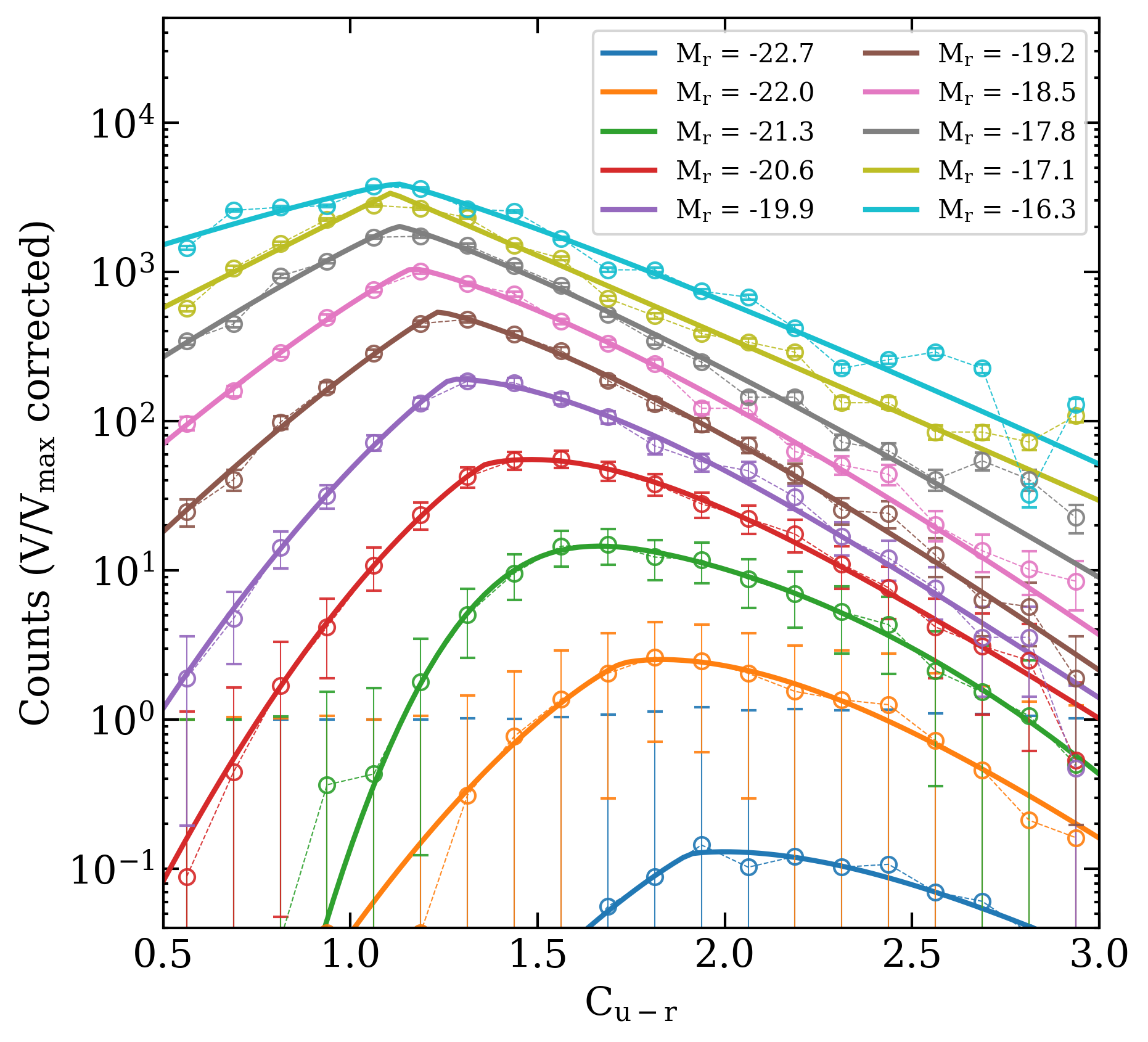}

\caption{The estimated r-band luminosity function from the A100-SDSS catalogue: The open circles represent the luminosity function computed using $1/V_{\max}$ method. Here, $V_{\max}$ is estimated considering both the \HI and optical flux limits, since the catalogue is flux-limited in both. The black solid line shows the best fit, with the corresponding $\chi^2$ values given in the legend, while the blue dashed line represents the $r$-band luminosity function from the SDSS optical catalogue as estimated in \citep{Baldry}. The \emph{right-panel} shows the distribution of number counts as a function of (u–r) colour. Different coloured points represent the number counts in various r-band magnitude bins, as indicated in the legend. The corresponding coloured solid curves are the best-fit models using a skewed Gaussian function. The error bars represent the $1\sigma$ Poisson uncertainties.}

\label{fig:A100SDSS_LF_CM}
\end{figure*}

The r-band luminosity function is estimated from the A100-SDSS observational data. As data is both \HI and optical (SDSS) selected here, we use \HI apparent flux and r-band apparent flux to compute the maximum volume i.e. $V_{\rm max} = min(V^{\rm Opt}_{\rm max},V^{\rm \HI}_{\rm max})$. Here $V^{\rm \HI}_{\max}$~($V^{\rm Opt}_{\max}$) denotes the maximum comoving volume of the Universe within which a galaxy of a given absolute \HI flux (r-band luminosity) can be detected by the survey, based on its apparent flux threshold. The joint luminosity function can be estimated using $1/V_{\rm max}$ method \citep{Schmidt}, which is given as,
    \begin{equation}
        \phi(M_r)\Delta M_r = \sum^N_i \frac{1}{V_{\rm max}}
        \label{L}
    \end{equation}
Where $i$ runs over all the galaxies within a magnitude bin. We fit the luminosity function using the double Schechter function~\citet{Schechter}, which can be given as,
\begin{eqnarray}
        \phi(M_r) &=& c\phi_1^*e^{-c(\alpha_1+1)(M_r-M^*)}e^{-e^{-c(M_r-M^*)}} \nonumber \\ 
        && + c\phi_2^*e^{-c(\alpha_2+1)(M_r-M^*)}e^{-e^{-c(M_r-M^*)}}
        \label{eq:double_sch}
\end{eqnarray}
where $c=0.4\ln(10)$; $M^*$~and~$\phi^*$ are characteristic   magnitude and number density and $\alpha$ is faint-end slope. Most of the galaxies in the A100–SDSS catalogue are star-forming (blue) \HI-selected systems. Therefore, we compare the luminosity function estimate with that of the blue galaxy population in the SDSS optical surveys, as reported by \citet{Baldry}. In  \emph{left-panel} of Fig.~\ref{fig:A100SDSS_LF_CM}, we present the estimate of luminosity function using $1/V_{\max}$ method. The black solid line shows the best fit, with the corresponding $\chi^2$ values given in the legend, while the blue dashed line represents the $r$-band luminosity function from the SDSS optical catalogue as reported in \citet{Baldry}. Our luminosity function estimate is consistent within $2\sigma$ confidence.

In the \emph{right-panel}, we plot the number count of  (u-r) colour for 10 magnitude bins. Applying the $V/V_{\max}$ correction ensures that the counts correspond to a volume-limited colour distribution in each magnitude bin. Where $V$ is the survey volume. The error bars represent the $1\sigma$ Poisson uncertainties. We fit a skewed Gaussian function to model the number count distribution in each magnitude bin.
\begin{equation}
f(x) \;=\; A\exp\!\left[-\left|\frac{x - \mu}{\sigma}\right|^{\beta}\right]
\, \Phi\!\left(\frac{\alpha (x-\mu)}{\sigma}\right)
\end{equation}
where $A$ is normalization factor; $\mu$ is mean of the symmetric component; $\sigma$ controls the width of the distribution; $\beta$ controls the sharpness or tail behaviour; $\alpha$ is skewness and $\Phi$ is standard normal cumulative distribution function. The motivation for using a skewed Gaussian is that the A100-SDSS galaxy sample is \HI-selected as the majority of red galaxies ( i.e. \HI gas-poor or quenched galaxies) are not detected in ALFALFA, which causes the typically bimodal Gaussian colour distribution to appear as a single-skewed distribution.

%%%%%%%%%%%%%%%%%%%%%%%%%%%%%%%%%%%%%%%%%%%%%%%%%%%%%%%%%%%%%%%%%%%%%%%%
%%%%%%%%%%%%%%%%%%%%%%%%%%%%%%%%%%%%%%%%%%%%%%%%%%%%%%%%%%%%%%%%%%%%%%%%

%%%%%%%%%%%%%%%%%%%%%%%%%%%%%%%%%%%%%%%%%%%%%%%%%%%%%%%%%%%%%%%%%%%%%%%%
\subsection{Relations: mass-redshift-inclination}
\label{ssec:relations}
\begin{figure*}
\centering
\includegraphics[width=.33\textheight]{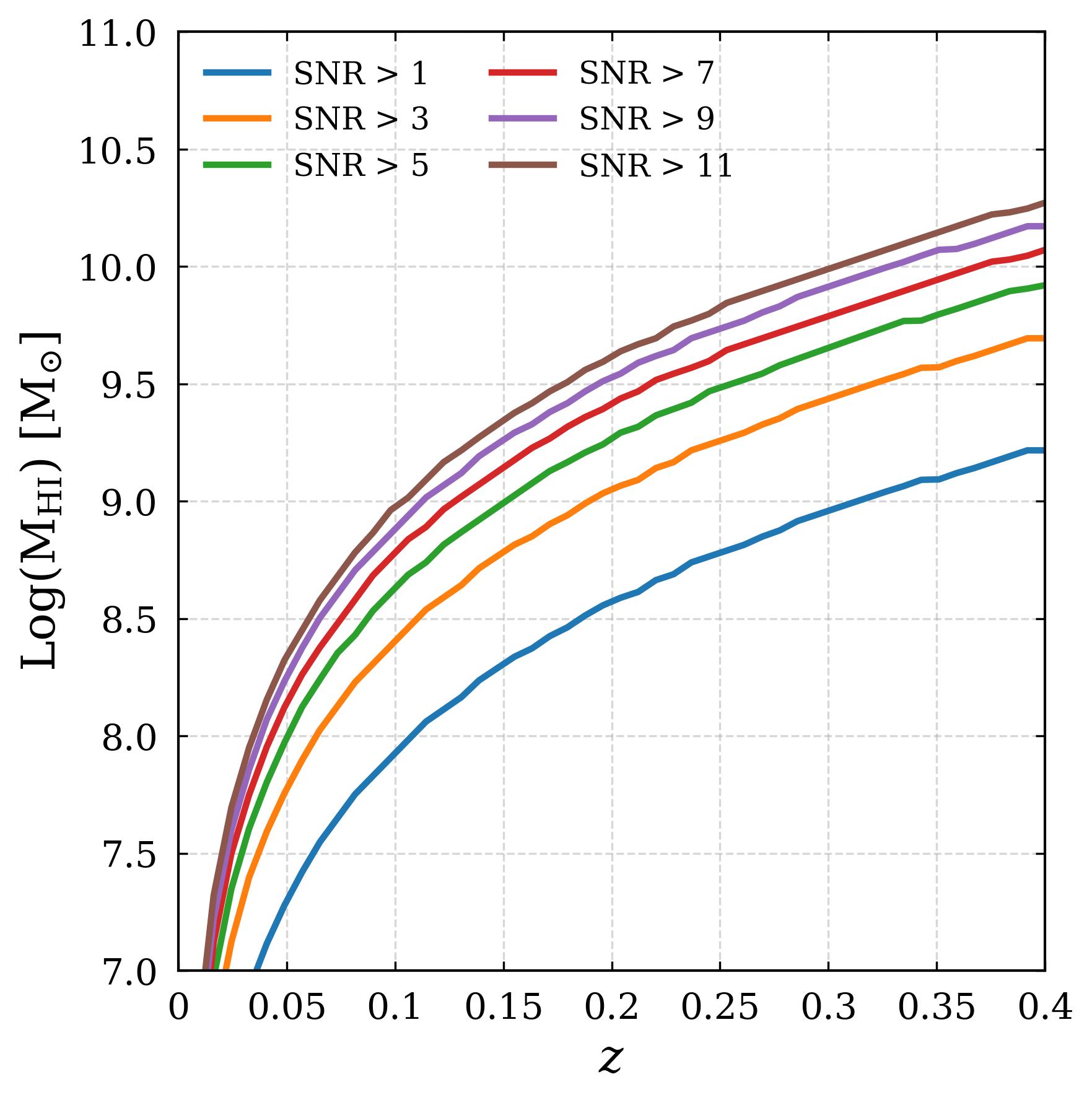}
\centering\includegraphics[width=.33\textheight]{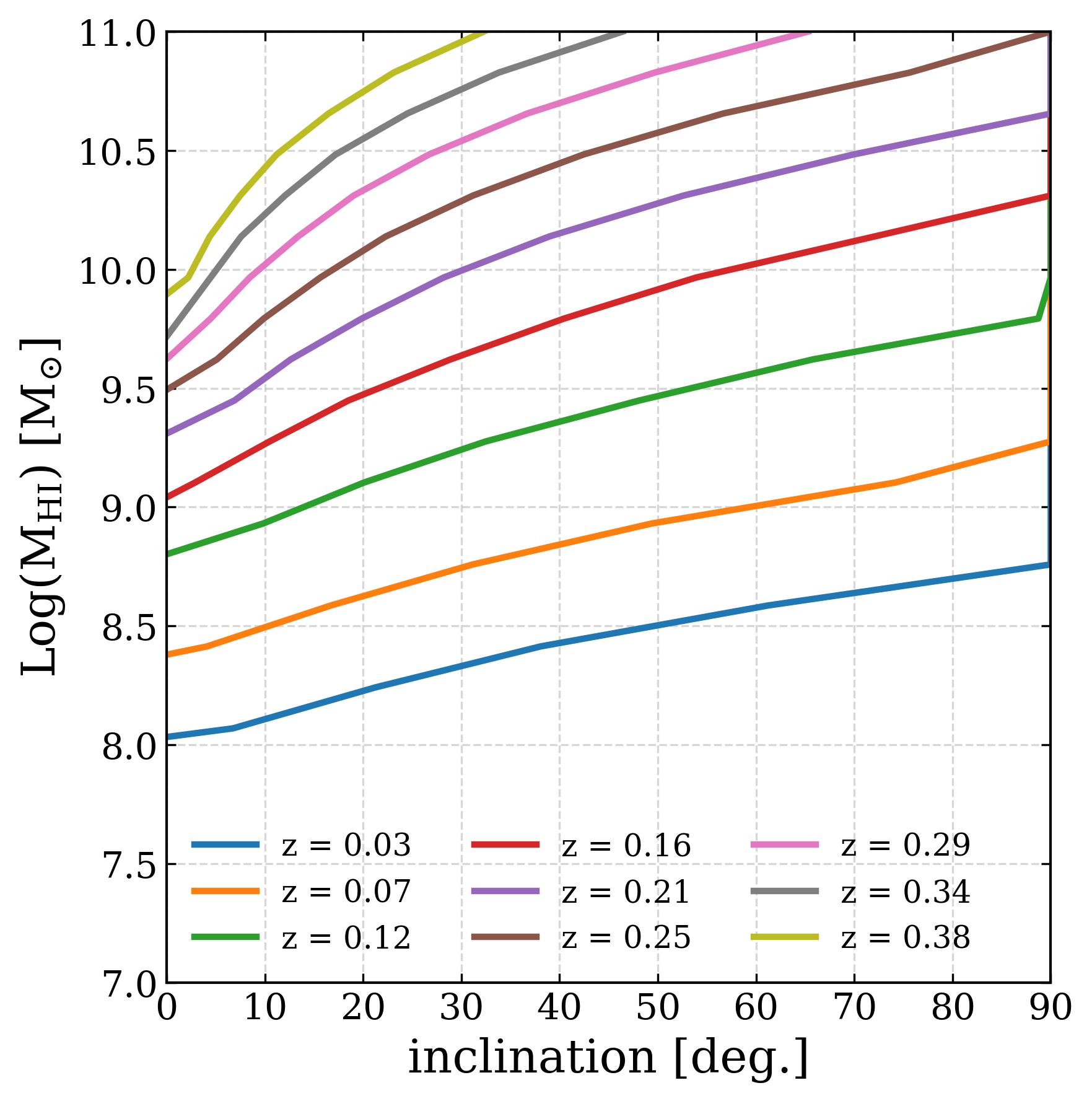}

\caption{\emph{First panel} shows mass-redshift limits used in triple integration at a given SNR threshold. The second panel shows the accessible range of the maximum possible inclination angle that a source can have with $SNR > 5$ in the MeerKAT mock survey.}\label{fig:mz_relation}
\end{figure*}

Given a detection threshold (say $\mathrm{SNR>5.0}$), we compute the mass-redshift relation defined by Eq.~\ref{eq:HIsignal}. The \HI source having mass $M_{\mathrm{\HI}}$ and redshift $z$ can be detected up to a maximum comoving volume at redshift $z_{\max} \ge z$, given the flux limit of the survey. These sources follow a simple relation in the mass--redshift plane for a given detection threshold, as shown in the \emph{left-panel} of Fig.~\ref{fig:mz_relation}. To compute the second integral in Eq.~\ref{eq:count_model}, we use these relations. Similarly, for a given mass and redshift, the inclination of the source also plays a key role in defining the detection threshold. For example, face-on sources (with sharper peak signals) are easier to detect compared to edge-on sources. Thus, for a fixed detection threshold, a source with specified redshift and \HI mass can be detected only up to a maximum inclination angle $I_m$. \emph{Right-panel} of Fig.~\ref{fig:mz_relation} shows how maximum inclination $I\le I_m$ angle vary as function of \HI mass and redshift of the source. These relations are for the source which can be detected with $\mathrm{SNR>5}$ and used in computing the third integral in Eq.~\ref{eq:count_model}. It is evident that sources at $z = 0.38$ with \HI masses in the range $M_{\mathrm{HI}} = 10^{10} - 10^{11}~M_{\odot}$ can only be detected for inclination angles $I \leq 30^\circ$. For inclinations $I > 30^\circ$, the signal falls below the detection threshold, or a higher \HI{} mass ($> 10^{11}~M_{\odot}$) would be required for detection.
%%%%%%%%%%%%%%%%%%%%%%%%%%%%%%%%%%%%%%%%%%%%%%%%%%%%%%%%%%%%%%%%%%%%%%%%

% Don't change these lines
\bsp	% typesetting comment
\label{lastpage}
\end{document}